\documentclass{article}
\usepackage[a4paper,
            bindingoffset=0.2in,
            left=0.6in,
            right=0.6in,
            top=1in,
            bottom=1in,
            footskip=.25in]{geometry}
\usepackage[utf8]{inputenc}
\usepackage{jcappub}
\usepackage{natbib}
\setcitestyle{square,numbers}
\usepackage{bbold}
\usepackage{graphicx}
\usepackage{floatpag}
\usepackage{xcolor}
\usepackage{float}
\usepackage{multirow,array,longtable}
\usepackage{arydshln} 
\usepackage[makeroom]{cancel}
\usepackage{amsfonts}
\usepackage{stackrel}

\renewcommand\eqref[1]{Eq.~(\ref{#1})}

\newcommand\figref[1]{Fig.~\ref{#1}}
\newcommand\figrefs[2]{Figs.~\ref{#1}-\ref{#2}}

\newcommand\secref[1]{Section~\ref{#1}}
\newcommand\appref[1]{Appendix~\ref{#1}}
\newcommand{\be}{\begin{equation}}
\newcommand{\ee}{\end{equation}}
\newcommand{\bear}{\begin{eqnarray}}
\newcommand{\eear}{\end{eqnarray}}
\newcommand{\nn}{\nonumber}

\newcommand{\mM}{\mathcal{M}}
\newcommand{\mH}{\mathcal{H}}
\newcommand{\mN}{\mathcal{N}}
\newcommand{\mS}{\mathcal{S}}
\newcommand{\mC}{\mathcal{C}}

\newcommand{\mI}{\mathcal{I}}
\newcommand{\mB}{\mathcal{B}}
\def\BOp{\hat{\mB}_{CGLMP}}
\def\vNS{\mS_{\rm EE}}

\def\mw{m_{\rm W}}

\def\wpm{W^\pm}
\def\ww{W^+W^-}
\def\Tr{{\rm Tr}}

\title{Exploring Bell inequalities and quantum entanglement in vector boson scattering}
\author[a]{R. A.  Morales}
\affiliation[a]{IFLP, CONICET - Dpto. de F\'{\i}sica, Universidad Nacional de La Plata, \\ 
C.C. 67, 1900 La Plata, Argentina}
\emailAdd{roberto.morales@fisica.unlp.edu.ar}

\abstract{
Quantum properties of vector boson scattering $V'_1V'_2\to V_1 V_2$, related to entanglement and violation of Bell inequalities, are explored in this paper.
The analysis is based on the construction of the polarization density matrix associated to the final state $V_1V_2$ by means of the computation of the corresponding tree level amplitudes within the Standard Model.
The aim of this work is to determine the regions of the phase space where the final vector bosons after the scattering result entangled and if is it possible to test the Bell inequalities in those regions.
We found that in all cases the entanglement is present.  
The amount of it depends on the process and the Maximally Entangled state is reached in some particular channels.
Concerning the Bell inequality, it could be also tested in certain kinematical regions for some of these processes.
This work is a first step in the analysis of these quantum properties for this kind of processes and it is postponed for future studies the reconstruction of the polarization density matrix and the related quantum parameters from experimental data through Monte-Carlo simulations using quantum tomography techniques.}

\begin{document}

\maketitle

\section{Introduction}
\label{section-intro}

Quantum entanglement plays a fundamental role in communication between particles separated by macroscopic distances~\cite{Horodecki:2009zz}. It is due to the presence of correlations in quantum systems which cannot be replicated by classical systems. 
One of the most relevant consequences is the violation of Bell inequality~\cite{PhysicsPhysiqueFizika.1.195},  a nonviable fact in any theory consistent with the classical concepts of realism and locality.
This special feature of quantum systems was the key for discovering quantum technologies as cryptography~\cite{PhysRevLett.67.661}, teleportation~\cite{PhysRevLett.70.1895} and quantum computation~\cite{PhysRevLett.86.5188}. 
Furthermore, theoretical physicists make remarkable progresses on this topic in the context of quantum field theories, see for instance the review~\cite{Casini:2022rlv}.

The study of quantum entanglement and Bell inequalities in the context of High Energy Physics, from both phenomenological and experimental point of views, has received a very recent attention.
In particular, it provides the possibility of investigate quantum entanglement in a relativistic limit with the highest possible energies at colliders.
Concretely, testing Bell inequalities have been proposed in $e^-e^+$ collisions~\cite{Tornqvist:1980af}, charmonium~\cite{Baranov_2008,10.1093/ptep/ptt032} and positronium~\cite{PhysRevA.63.042107} decays, neutrino oscillations~\cite{Blasone_2009,Banerjee_2015} and more recently for Higgs decays into gauge bosons~\cite{Barr:2021zcp,Aguilar-Saavedra:2022mpg,Aguilar-Saavedra:2022wam,Ashby-Pickering:2022umy,Fabbrichesi:2023cev}.
Also proposals to test Bell inequalities have been made in systems composed by mesons~\cite{Bertlmann_2001,Banerjee_2016,Takubo_2021,fabbrichesi2023bell}, top-quarks~\cite{Afik:2020onf,Fabbrichesi:2021npl,Severi:2021cnj,Afik:2022kwm,Aoude:2022imd,Aguilar-Saavedra:2022uye,Severi:2022qjy,Dong:2023xiw}, tau-lepton and photon pairs~\cite{Fabbrichesi:2022ovb,Altakach:2022ywa} and two massive gauge bosons~\cite{Barr:2022wyq,Fabbrichesi:2023cev}.

Violation of Bell inequalities has entanglement as a necessary but not sufficient condition, then both concepts deserve dedicated attention.
In particular, the violation of such inequalities probes that there is no hidden variable theory for accomplished the generated entanglement.
The aim of this work is to perform a systematic study of them through vector boson scattering (VBS) processes in the SM, following closely~\cite{Aguilar-Saavedra:2022wam,Ashby-Pickering:2022umy,Fabbrichesi:2023cev,Fedida:2022izl}.
The latter of these previous works has presented a complete theoretical analysis of entanglement in QED processes. The others were focused on testing both entanglement and Bell inequalities for Higgs boson decays into massive gauge bosons and diboson production at the LHC and future lepton colliders by means of Monte-Carlo simulations.
In particular, a general method for the experimental reconstruction of the density matrix and related entanglement measurements from angular decay data was developed in~\cite{Ashby-Pickering:2022umy}. This technique is known as quantum tomography and, in the context of High Energy Physics, was previously applied to lower-dimension systems, see for instance the $t\bar{t}$ case~\cite{Afik:2020onf}. 
Now the VBS processes allow to study a variety of bipartite systems with more intricate density matrices compared to those studied in $t\bar{t}$-pair and Higgs boson decays.

Undoubtedly, the relevance of the VBS processes for examining the deepest structure of the electroweak (EW) interactions in the SM is well established in the community. 
In particular, the precise cancellation of potentially large contributions among diagrams with only trilinear gauge self-couplings, quartic gauge self-couplings and with the Higgs boson, is responsible for unitarity restoration in this kind of processes.
Then probing these interactions reveal the dynamics behind the Higgs mechanism. In this line, there is an active program of experimental searches by ATLAS and CMS Collaborations, see for instance the recent reviews~\cite{Covarelli:2021gyz,BuarqueFranzosi:2021wrv}.
Moreover, VBS is a suitable observable for new physics in the EW sector looking for anomalous triple (aTGC) and quartic (aQGC) gauge couplings and also with Higgs boson interactions.
Taking these considerations into account, the goal of this paper is to perform theoretical predictions of entanglement and violation of Bell inequalities for different VBS processes, which are not explored in the literature as far as we know. 
In particular, this analysis allows us to locate kinematical regions of the phase space where interesting quantum mechanical measurements might be performed.
The determination of the related quantities from Monte-Carlo simulations in the complete collider events, i.e. a quantum tomography analysis, is beyond the scope of this work and it is postponed for a future study.
\newline

The paper is organized as follows: 
\secref{section-methods} presents the theoretical framework connecting the scattering amplitudes and the density matrix formalism's. The quantifiers related to entanglement and Bell inequalities are also introduced in this section.
The main results of this work for the VBS processes are collected in \secref{section-numresults}. This section is closed with a brief discussion of a possible quantum tomography analysis for this kind of processes at colliders.
We summarize the main findings and future perspectives in \secref{section-conclus}.
The appendices contain the details for the amplitude computation, additional plots and analytical expressions for $\wpm\gamma\to\wpm\gamma$ and $\gamma\gamma\to\ww$ processes.

\section{Formalism}
\label{section-methods}

The traditional approach to QFT in Particle Physics focuses on scattering amplitudes by means of Feynman diagrams. In this work, we will consider 2$\to$2 particle scattering processes among vector bosons within the SM,
\be
V'_1(p'_1,s'_1)+V'_2(p'_2,s'_2) \to V_1(p_1,s_1)+V_2(p_2,s_2)
\label{eq-VBSproc}
\ee
where $V_i$ ($V'_i$) denotes final (initial) photons, $\wpm$ or $Z$ bosons with momentum $p_i$ ($p'_i$) and polarization $s_i$ ($s'_i$), for $i=1,2$. In particular, `$\pm$' denote the transverse polarizations and `$0$' is used for the longitudinal one (only for the massive gauge bosons). 

In the language of Quantum Information Theory (QIT), knowledge about the quantum system is represented by the density matrix $\rho$.
Seminal works~\cite{Fano:1957zz,rhoStanton,Hagston_1980} relate the density matrix formalism with the S-matrix operator for the scattering processes.
In this work, we will focus on the quantum entanglement of the vector boson polarizations $s_i$.
The bipartite final state system $\vert f\rangle$ is defined in the Hilbert space $\mH_1\otimes\mH_2$ where each $\mH_{i}$ has dimension $d_i=dim(\mH_i)$ equals to 2 for photons or to 3 for massive $\wpm$ and $Z$ gauge bosons. 
In QIT language, photons correspond to qubits and massive gauge bosons are qutrits.
In addition, the chosen basis corresponds to the spin-eigenvalue assignment for the outcomes $\{+,-\}$ and $\{+,0,-\}$, respectively.
In this sense, we write $\vert f\rangle=\vert s_1\rangle\otimes\vert s_2\rangle$ where momenta is omitted for simplicity. Similar definitions for the initial state $\vert i\rangle=\vert s'_1\rangle\otimes\vert s'_2\rangle$ in the Hilbert space $\mH'_1\otimes\mH'_2$ are implemented.
The unitary time evolution of two incoming particles state $\vert i\rangle$ into two outgoing particles state $\vert f\rangle$ is given by the scattering amplitude $\mM(V'_1V'_2\to V_1V_2)$ which is related to the S-matrix elements by momentum conservation relation $S_{fi}=i(2\pi)^4\delta^{(4)}(p'_1+p'_2-p_1-p_2)\mM(V'_1V'_2\to V_1V_2)$.
For the present calculation, these amplitudes were computed in the SM at tree level and relevant details on the kinematics are summarized in the \appref{App-kinem}.

The density matrix at $t_0=-\infty$ for unpolarized initial state is
\be
\rho(t_0=-\infty) = \vert i\rangle\langle i\vert  = \sum_{s'_1,\,s'_2} \frac{1}{d'_1d'_2}\vert s'_1\,s'_2\rangle\langle s'_1\,s'_2\vert
\ee
where $d'_i=dim(\mH'_i)$ and $\vert s'_1\,s'_2\rangle=\vert s'_1\rangle\otimes\vert s'_2\rangle$ for a compact notation.
The unitary time-evolution of this density matrix is performed through the S-matrix operator~\cite{Hagston_1980,Fedida:2022izl} and the resulting filtered state is
\be
\rho(t=+\infty) = \frac{1}{N}\sum_{f,\tilde{f}} \vert f\rangle\langle f\vert S\rho(t_0=-\infty)S^\dagger \vert \tilde{f}\rangle\langle \tilde{f}\vert = \frac{1}{N}\sum_{f,\tilde{f}} \vert f\rangle S_{fi} S_{\tilde{f}i}^* \langle \tilde{f}\vert
\label{rho-out}
\ee
where the normalization factor $N$ is fixed by the condition Tr$[\rho(t=+\infty)]=1$. From now on, we will write $\rho$ instead of $\rho(t=+\infty)$ since entanglement of the final state polarizations is computed here.

Finally, the corresponding bipartite density matrix elements are
\be
\langle s_1\,s_2\vert \rho \vert \tilde{s}_1\,\tilde{s}_2\rangle =\frac{\mM_{s_1,s_2}\mM^\dagger_{\tilde{s}_1,\tilde{s}_2}}{\vert\overline{\mM}\vert^2}
\label{rhospin}
\ee
where $\mM_{s_1,s_2}$ is the averaged amplitude $\mM(V'_1V'_2\to V_1V_2)$ over all initial state polarizations $s'_i$. Also, $\vert\overline{\mM}\vert^2$ is the total unpolarized square amplitude summing over final state which accounts for the normalization factor $N$ in \eqref{rho-out}.
It is important to stress that these amplitudes depends on two kinematical variables, the energy and the scattering angle, then the $\rho$ matrix does too.

In order to reconstruct the density matrix of each scattering process from experimental data, i.e. the goal of quantum tomography, it is useful to introduce the following parametrization~\cite{Ashby-Pickering:2022umy} 
\be
\rho = \frac{1}{d_1d_2}I_{d_1d_2} +\frac{1}{2d_2}\sum_{i=1}^{d_1^2-1}A_i\lambda_i^{(d_1)}\otimes I_{d_2} +\frac{1}{2d_1}\sum_{j=1}^{d_2^2-1}B_j I_{d_1}\otimes\lambda_j^{(d_2)} +\frac{1}{4}\sum_{i=1}^{d_1^2-1}\sum_{j=1}^{d_2^2-1}C_{ij} \lambda_i^{(d_1)}\otimes\lambda_j^{(d_2)} 
\label{rho-decom}
\ee
where $I_{n}$ is the dimension-$n$ identity matrix and the $\lambda^{(d)}$'s are the dimension-$d$ generalized Gell-Mann matrices.
The relevant ones for this work are the three Pauli matrices $\vec{\sigma}=\left(\lambda^{(2)}_1,\lambda^{(2)}_2,\lambda^{(2)}_3\right)$ for the qubits and the eight Gell-Mann matrices $\vec{\lambda}=\left(\lambda^{(3)}_1,...\,,\lambda^{(3)}_8\right)$ for the qutrits (see \appref{App-matrices}).
The coefficients $A_i={\rm Tr}\big[\rho\cdot\lambda_i^{(d_1)}\otimes I_{d_2}\big]$ and $B_j={\rm Tr}\big[\rho\cdot I_{d_1}\otimes\lambda_j^{(d_2)}\big]$ provide the polarization of the gauge bosons. Moreover, the matrix $C_{ij}={\rm Tr}\big[\rho\cdot \lambda_i^{(d_1)}\otimes\lambda_j^{(d_2)}\big]$ represents the correlation among them. 

\subsection{Entanglement vs separability}

The knowledge of the density matrix allows the computation of different entanglement quantifiers in order to determine the level of the correlations in the corresponding system.
Furthermore, since entanglement is a necessary but not sufficient condition for the violation of Bell inequalities, it is pertinent to test both phenomena in an energy regime never explored before.
In particular, both concepts impose restrictions on $C_{ij}$, i.e. assess particular relations between coefficients of the correlation matrix.
A `separable' bipartite state can be expressed as 
\be
\rho_{\rm sep} = \sum_n p_n\rho_n^{(V_1)}\otimes\rho_n^{(V_2)}
\label{rhosep}
\ee
where $\rho_n^{(V_1)}$ and $\rho_n^{(V_2)}$ are the density matrices of the subsystems $V_1$ and $V_2$, respectively, and $p_n$ are classical probabilities adding 1. 
These separable states can be created by classical communications and local operations. On the contrary, a bipartite state is defined as `entangled' if it is not separable, then its density matrix cannot be written as in \eqref{rhosep}.
In practice it is hard to decide if a given state is separable or entangled based on the previous deﬁnition and this is called the `separability problem'.
There are no known necessary and sufficient conditions for evaluating quantum entanglement of a bipartite system with arbitrary dimension~\cite{Horodecki:2009zz}.
The VBS processes allow us to explore different cases corresponding to two-qubits ($2\otimes 2$ dimension), qutrit$\otimes$qubit ($3\otimes 2$) and two-qutrits ($3\otimes 3$). 

For two-qubits and qutrit$\otimes$qubit cases, the Positivity of the Partial Transpose (PPT), also called Peres-Horodecki criterion, gives the necessary and sufficient conditions for entanglement~\cite{Peres:1996dw,Horodecki:1996nc}. Concretely, the partially transpose matrix given by
\be
\langle s_1\,s_2\vert \rho^{\rm T_2} \vert\tilde{s}_1\,\tilde{s}_2\rangle = \langle s_1\,\tilde{s}_2\vert \rho \vert\tilde{s}_1\,s_2\rangle 
\ee
has at least one negative eigenvalue if and only if the density matrix in \eqref{rhospin} represents an entangled state. Denoting by $\lambda^{\rm T_2}_k$ the eigenvalues of $\rho^{\rm T_2}$, the amount of the system entanglement is computed by the Negativity~\cite{PhysRevA.65.032314}
\be
\mN(\rho) = \sum_{k} \frac{\vert\lambda^{\rm T_2}_k\vert-\lambda^{\rm T_2}_k}{2}
\label{definition-Neg}
\ee
Notice that the Negativity vanishes if and only if all the eigenvalues of the partially transpose density matrix $\rho^{\rm T_2}$ are non-negative.

The PPT criterion is just a sufficient condition for entanglement in the two-qutrits case. Only in some special cases it is also a necessary condition. For example, it is shown in~\cite{Aguilar-Saavedra:2022wam,Aguilar-Saavedra:2022mpg} that PPT criterion is also a necessary condition for the Higgs boson decays into $ZZ$ and $\ww$. 
Furthermore, the general two-qutrits processes considered in this work lead to density matrices in \eqref{rhospin} which correspond to pure states, i.e. $\rho$ is idempotent and acts as a projector. 
For these bipartite pure states, the Entropy of Entanglement $\vNS$~\cite{PhysRevA.53.2046}, that is the von Neumann Entropy of either subsystem $V_1$ or $V_2$, can be computed as
\bear
\vNS(\rho) &=& -{\rm Tr}[\rho_{red\,1}\log\rho_{red\,1}] = -{\rm Tr}[\rho_{red\,2}\log\rho_{red\,2}]  \nn\\
&=& -\sum_l \lambda^{red}_l\log(\lambda^{red}_l)
\label{definition-vNS}
\eear
where the partial trace over each subsystem yields to the reduced density matrices $\rho_{red\,1}={\rm Tr}_2[\rho]=\sum_{s_2}\langle s_2\vert\rho\vert s_2\rangle$ and $\rho_{red\,2}={\rm Tr}_1[\rho]=\sum_{s_1}\langle s_1\vert\rho\vert s_1\rangle$. Both reduced matrices have the same non-vanishing eigenvalues $\lambda^{red}_l$ and the entropy of entanglement can be computed by the sum over the non-zero eigenvalues of any reduced matrix, as in the second line of the previous equation.
The Entropy of Entanglement vanishes if and only if the pure state is separable.

Another relevant entanglement quantifier is the `Concurrence' of the system, which for a two-qutrits pure state is computed as~\cite{Rungta_2001}
\be
\mC[\rho] = \sqrt{2(1-\Tr[(\rho_{red\,1})^2])} = \sqrt{2(1-\Tr[(\rho_{red\,2})^2])}
\label{definition-Concu}
\ee
which is a generalization of the two-qubit Concurrence case~\cite{Hill:1997pfa}.
Therefore the corresponding state is separable if and only if the Concurrence is equal to zero. 

The Maximally Entangled two-qubits ($d=2$) and two-qutrits ($d=3$) states are defined as
\be
\vert\Psi_{\rm MaxEnt}\rangle = \frac{1}{\sqrt{d}}\sum_{k=1}^d \vert k\rangle\otimes\vert k\rangle
\ee
where $\vert k\rangle$ correspond to the orthonormal basis of the Hilbert spaces $\mH_{1,2}$. 
For these particular states, the Negativity, Entropy of Entanglement and Concurrence reach their theoretical maximum values $\frac{d-1}{2}$, $\log(d)$ and $\sqrt{\frac{2(d-1)}{d}}$, respectively~\cite{Eltschka_2015}.

\subsection{Testing Bell inequalities}

Besides the entanglement due to correlations in a quantum system, a stronger requirement is the violation of Bell inequalities.
The quantifier $\mI$ is introduced in order to discriminate among predictions coming from deterministic local theories from those coming from quantum mechanics. In particular, local realist models must obey the so-called Bell inequality
\be
\mI \leq 2
\label{Bellineq}
\ee
which can be violated in QFT.
For the two-qubits cases, the optimal Clauser-Horne-Shimony-Holt (CHSH) operator~\cite{CHSH} is defined as
\be
\mB_{CHSH} = \hat{a}_1\otimes\hat{b}_1 -\hat{a}_1\otimes\hat{b}_2 +\hat{a}_2\otimes\hat{b}_1 +\hat{a}_2\otimes\hat{b}_2
\label{eq-BellOP-CHSH}
\ee
with $\hat{a}_i=\vec{a}_i\cdot\vec{\sigma}$ and $\hat{b}_i=\vec{b}_i\cdot\vec{\sigma}$ are Hermitian operators acting on the Hilbert spaces $\mH_1$ and $\mH_2$, respectively.
In practical computations, the quantifier $\mI_{2\otimes 2}\equiv\mI_2$ is defined as the maximal expectation value of this Bell operator over the unit vectors $\vec{a}_i$ and $\vec{b}_i$ in $\mathbb R^3$
\bear
\mI_2 &=& \stackbin[\vec{a}_i,\vec{b}_i]{}{\rm Max}\big\{\Tr[\rho\cdot\mB_{CHSH}]\big\}  \nn\\
&=& \stackbin[\vec{a}_i,\vec{b}_i]{}{\rm Max}\big\{\vec{a}_1^{\rm T}C(\vec{b}_1-\vec{b}_2)+\vec{a}_2^{\rm T}C(\vec{b}_1+\vec{b}_2)\big\}  \nn\\
&=& 2\sqrt{r_1+r_2}
\label{definition-I2}
\eear
where the decomposition of the density matrix in \eqref{rho-decom} was used in the second line (notice that only the correlation matrix $C$ is relevant for this expectation value).
In the third line, $r_1$ and $r_2$ are the two largest eigenvalues of the matrix $C^{\rm T}C$~\cite{HORODECKI1995340}. Hence the Bell inequality in \eqref{Bellineq} can be violated if and only if $r_1+r_2$ is larger than 1. Moreover, the maximum theoretical value is the Cirelson bound $2\sqrt{2}$~\cite{Cirelson:1980ry}, corresponding to the Maximally Entangled state.

For the qutrit$\otimes$qubit cases, and in a similar way than in~\cite{Caban:2008qa}, we consider a generalization to the previous CHSH. Now the 3$\otimes$2 Bell operator is 
\be
\mB_{CHSH}^{gen} = \vec{n}_1\cdot\vec{S}\otimes(\vec{n}_2-\vec{n}_4)\cdot \vec{\sigma} +\vec{n}_3\cdot\vec{S}\otimes(\vec{n}_2+\vec{n}_4)\cdot \vec{\sigma} 
\label{eq-BellOP-genCHSH}
\ee
where $\vec{n}_i$ are unit vectors in $\mathbb R^3$ and the dimension-3 spin-1 matrices $\vec{S}=(S_x,S_y,S_z)$ are shown in \eqref{eq-spinmatrices}. The resulting quantifier is
\bear
\mI_{3\otimes 2} &=& \stackbin[\vec{n}_i]{}{\rm Max}\big\{\Tr[\rho\cdot\mB_{CHSH}^{gen}] \big\}  \nn\\
&=& 2\sqrt{\tilde{r}_1+\tilde{r}_2}
\label{definition-I3x2}
\eear
where, following the two-qubits case, $\tilde{r}_{1,2}$ are the two largest eigenvalues of $\tilde{C}^{\rm T}\tilde{C}$.
In this case, the spin correlation matrix $\tilde{C}$ is defined from the decomposition of the density matrix in \eqref{rho-decom} as
\be
\tilde{C}_{1j}=\frac{1}{\sqrt{2}}(C_{1j}+C_{6j})\,, \qquad \tilde{C}_{2j}=\frac{1}{\sqrt{2}}(C_{2j}+C_{7j})\,, \qquad \tilde{C}_{3j}=\frac{1}{2}(C_{3j}+\sqrt{3}C_{8j})
\ee

For the two-qutrits cases, the optimal quantifier $\mI_{3\otimes 3}\equiv\mI_3$ correspond to Collins-Gisin-Linden-Massar-Popescu (CGLMP)~\cite{CGLMP}.
This $\mI_3$ is constructed as the expectation value of an appropriate Bell operator $\BOp$ as usual, $\Tr[\rho\cdot\BOp]$.
A clever election of the $\BOp$ operator improves the violation of Bell inequality in \eqref{Bellineq}.
It was shown in~\cite{Barr:2021zcp} that a suitable operator for Higgs boson decaying into gauge boson pair corresponds to 
\be
\BOp^{xy} = -\frac{2}{\sqrt{3}}\left(S_x\otimes S_x +S_y\otimes S_y\right) +\lambda_4\otimes\lambda_4 +\lambda_5\otimes\lambda_5
\label{eq-CGLMPxy}
\ee
where the spin and the Gell-Mann matrices in dimension-3 are collected in the \appref{App-matrices}. 
In turn, local unitary changes of basis of the states $\vert s_1\rangle\otimes\vert s_2\rangle$, which define the density matrix, allow an optimization of the $\mI_3$ values for the VBS processes considered here. In particular, a maximization procedure for each $\rho$ density matrix is performed 
\be
\mI_3 = \stackbin[\vec{\alpha}_i,\vec{\beta}_i]{}{\rm Max}\big\{ \Tr[\rho\cdot(U_1\otimes U_2)^\dagger\cdot\BOp^{xy}\cdot(U_1\otimes U_2)] \big\}
\label{definition-I3}
\ee
where the maximization is over the domain $0\leq\alpha_1,\,\alpha_2\leq2\pi$ and $0\leq\beta_1,\,\beta_2\leq\pi$ for the angles defining the dimension-3 rotation matrices
\be
U_k = \exp(-iS_z\alpha_k)\exp(-iS_y\beta_k)
\ee
The maximum for $\mI_3$ is $1+\sqrt{11/3}\approx\,2.915$~\cite{PhysRevA.65.052325} and notoriously it is not achieved for the Maximally Entangled state which has $\mI_3^{\rm MaxEnt}=4(6\sqrt{3}+9)/27\approx\,2.873$.
\newline

In summary, the quantifiers related to entanglement detection and test Bell inequalities require the full knowledge of the quantum state which is collected in the density matrix. This $\rho$ matrix is determined using \eqref{rhospin} by computing the scattering amplitudes for the corresponding VBS process.
With this analysis, we can explore kinematical regions relevant for quantum mechanical measurements at colliders.
Depending on the final state, the following quantifiers will be derived,
\begin{itemize}
    \item for the two-qubits case: the only VBS process of this kind is $\ww\to\gamma\gamma$ and its Negativity $\mN$ of \eqref{definition-Neg}, the Entropy of Entanglement $\vNS$ of \eqref{definition-vNS} and $\mI_2$ of \eqref{definition-I2} will be determined.
    \item for the qutrit$\otimes$qubit case: also $\mN$, $\vNS$, and $\mI_{3\otimes2}$ of \eqref{definition-I3x2} will be calculated. The considered VBS processes of this kind are $\wpm\gamma\to\wpm\gamma$, $\ww\to Z\gamma$ and $\wpm Z\to\wpm\gamma$.
    \item for the two-qutrits case: the Entropy of Entanglement $\vNS$, Concurrence $\mC$ and $\mI_3$ of \eqref{definition-vNS}, \eqref{definition-Concu} and \eqref{definition-I3} will be computed, respectively. In this case, the considered VBS processes are $\gamma\gamma\to\ww$, $\wpm\gamma\to\wpm Z$, $\ww\to\ww$, $\wpm\wpm\to\wpm\wpm$, $\ww\to ZZ$, $\wpm Z\to\wpm Z$, $Z\gamma\to\ww$, $ZZ\to\ww$ and $ZZ\to ZZ$.
\end{itemize}

\section{Numerical Results}
\label{section-numresults}

The SM $2\to 2$ scattering amplitudes are computed at tree level in the center-of-mass frame (see \appref{App-kinem} for details) using FeynArts~\cite{FeynArts} and FormCalc~\cite{FormCalc-LT}. 
Some comments are in order: firstly, the radiative corrections can be relevant near the extreme theoretical values of the entanglement quantifiers in order to decide if the final state is separable or maximally entangled. Also for testing Bell inequality when $\mI\sim2$, however the generation of entanglement at the lowest order in perturbation theory is interesting by itself and we work on it.
Secondly, the amplitudes depend on two kinematical quantities corresponding to the scattering angle $\theta$ and the Mandelstam variable $S$ (related to the invariant mass of the final gauge bosons).
The numerical predictions of the mentioned entanglement quantifiers were computed in the plane $[\cos(\theta),\sqrt{S}]$ in the range -1 to 1 for the cosine function and from the scattering threshold up to 3 TeV for the energy.

\subsection{two-qubits case}

The only tree level VBS with photon-pair in the final state is $\ww\to\gamma\gamma$.
This lowest dimension case can be treated analytically and clarify the density matrix formalism from scattering amplitudes introduced in the previous section.
For a given photon-pair polarization $\vert s_1\,s_2\rangle$, the averaged amplitude over the initial state polarizations is
\bear
\mM_{s_1,s_2} &=& \frac{1}{9}\left (\mM(W^+_+W^-_+ \to\gamma_{s_1}\gamma_{s_2}) + \mM(W^+_+W^-_L \to\gamma_{s_1}\gamma_{s_2}) + \mM(W^+_+W^-_- \to\gamma_{s_1}\gamma_{s_2}) \right.  \nn\\
&& \left. +\mM(W^+_LW^-_+ \to\gamma_{s_1}\gamma_{s_2}) + \mM(W^+_LW^-_L \to\gamma_{s_1}\gamma_{s_2}) + \mM(W^+_LW^-_- \to\gamma_{s_1}\gamma_{s_2}) \right.  \nn\\
&& \left. +\mM(W^+_-W^-_+ \to\gamma_{s_1}\gamma_{s_2}) + \mM(W^+_-W^-_L \to\gamma_{s_1}\gamma_{s_2}) + \mM(W^+_-W^-_- \to\gamma_{s_1}\gamma_{s_2}) \right)  
\eear
where the sub-indices stand for the polarizations. The explicit computation of these amplitudes in terms of $S$ and $c=\cos(\theta)$ result in
\bear
\mM_{\pm,\pm} &=& \frac{1}{9}\frac{2e^2}{S(1-c^2)+4c^2\mw^2}\left( -12\left(1-c^2\right)\mw^2 +8\sqrt{2}ic\sqrt{1-c^2}\mw\sqrt{S} +\left(1+3c^2\right)S \right)  \nn\\
\mM_{\pm,\mp} &=& \frac{1}{9}\frac{8e^2}{S(1-c^2)+4c^2\mw^2}\left( S-3\mw^2 \right)
\eear
and the corresponding total unpolarized square amplitude is
\bear
\vert\overline{\mM}\vert^2 &=& \vert\mM_{++}\vert^2 +\vert\mM_{+-}\vert^2 +\vert\mM_{-+}\vert^2 +\vert\mM_{--}\vert^2  \nn\\
&=&\left(\frac{1}{9}\frac{2e^2}{S(1-c^2)+4c^2\mw^2}\right)^2 D_{\gamma\gamma}
\eear
with
\be
D_{\gamma\gamma} = 2 \left(144 \left(c^4-2 c^2+2\right) \mw^4-8 \left(7 c^4-10 c^2+15\right) \mw^2 S+\left(9 c^4+6 c^2+17\right) S^2\right)
\ee

Therefore in the basis $\{\vert++\rangle,\vert+-\rangle,\vert-+\rangle,\vert--\rangle\}$, the density matrix $\rho$ and its partially transpose $\rho^{\rm T_2}$ have a compact form
\be
\rho = \begin{pmatrix}
\rho_{11} & \rho_{12} & \rho_{12} & \rho_{11}  \\
\rho_{12}^* & \rho_{22} & \rho_{22} & \rho_{12}^*  \\
\rho_{12}^* & \rho_{22} & \rho_{22} & \rho_{12}^*  \\
\rho_{11} & \rho_{12} & \rho_{12} & \rho_{11}  \\
\end{pmatrix}
\qquad{\rm and}\qquad
\rho^{\rm T_2} = \begin{pmatrix}
\rho_{11} & \rho_{12}^* & \rho_{12} & \rho_{22}  \\
\rho_{12} & \rho_{22} & \rho_{11} & \rho_{12}^*  \\
\rho_{12}^* & \rho_{11} & \rho_{22} & \rho_{12}  \\
\rho_{22} & \rho_{12} & \rho_{12}^* & \rho_{11}  \\
\end{pmatrix}
\label{WWtoAA-rho}
\ee
where the three independent entries are:
\bear
\rho_{11} &=& \frac{1}{D_{\gamma\gamma}}\left( 144 \left(1-c^2\right)^2\mw^4 -8\left(7 c^4-10 c^2+3\right)\mw^2S +\left(1+3c^2\right)^2S^2 \right)  \nn\\
\rho_{12} &=& \frac{4(S-3\mw^2)}{D_{\gamma\gamma}}\left( -12\left(1-c^2\right) \mw^2 +8i\sqrt{2}c\sqrt{1-c^2}\mw\sqrt{S} + (1+3c^2)S \right)  \nn\\
\rho_{22} &=& \frac{16}{D_{\gamma\gamma}}\left( S-3\mw^2 \right)^2  
\label{WWtoAA-rhoindep}
\eear

The eigenvalues of the compact $\rho^{\rm T_2}$ in \eqref{WWtoAA-rho} can be written as
\bear
\lambda^{\rm T_2}_1 &=& \rho_{11}+\rho_{22}-\rho_{12}-\rho_{12}^*\,,  \qquad
\lambda^{\rm T_2}_2 = \rho_{11}+\rho_{22}+\rho_{12}+\rho_{12}^*\,,  \nn\\
\lambda^{\rm T_2}_3 &=& -\lambda^{\rm T_2}_4 = -\sqrt{(\rho_{11}-\rho_{22})^2-(\rho_{12}-\rho_{12}^*)^2}  
\eear

Using \eqref{WWtoAA-rhoindep}, the previous expressions for the $\ww\to\gamma\gamma$ scattering are 
\be
\lambda^{\rm T_2}_1 = \frac{U_1^{\gamma\gamma}}{D_{\gamma\gamma}}\,,\quad
\lambda^{\rm T_2}_2 = \frac{U_2^{\gamma\gamma}}{D_{\gamma\gamma}}\,,\quad
\lambda^{\rm T_2}_3 = -\frac{\sqrt{U_1^{\gamma\gamma}U_2^{\gamma\gamma}}}{D_{\gamma\gamma}}\,,\quad
\lambda^{\rm T_2}_4 = \frac{\sqrt{U_1^{\gamma\gamma}U_2^{\gamma\gamma}}}{D_{\gamma\gamma}}  \nn\\
\ee
in terms of the functions
\bear
U_1^{\gamma\gamma} &=&  144 c^4 \mw^4+56 c^2 \left(1-c^2\right) \mw^2 S+9 \left(1-c^2\right)^2 S^2  \nn\\
U_2^{\gamma\gamma} &=& 144 (2 - c^2)^2 \mw^4 - 
 8 (30 - 13 c^2 + 7 c^4) \mw^2 S + (5 + 3 c^2)^2 S^2 
\eear
Notice that $U_1^{\gamma\gamma}+U_2^{\gamma\gamma}=D_{\gamma\gamma}$ since Tr[$\rho^{\rm T_2}$]=Tr[$\rho$]=1.

In the relevant phase space of $[\cos(\theta),\sqrt{S}]$, the functions $D_{\gamma\gamma}$, $U_1^{\gamma\gamma}$ and $U_2^{\gamma\gamma}$ are positives. Therefore only the third eigenvalue $\lambda^{\rm T_2}_3$ is negative and the rest are positives. 
In consequence the Negativity is analytically given by
\be
\mN_{\ww\to\gamma\gamma} = \frac{\sqrt{U_1^{\gamma\gamma}U_2^{\gamma\gamma}}}{U_1^{\gamma\gamma}+U_2^{\gamma\gamma}}
\label{WWtoAA-Neg}
\ee

\begin{figure}
   \centering
  \begin{tabular}{cc}
\includegraphics[width=0.46\textwidth]{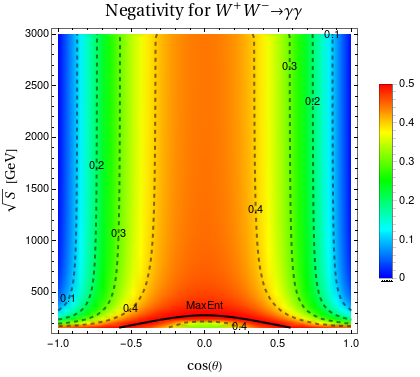} & \includegraphics[width=0.46\textwidth]{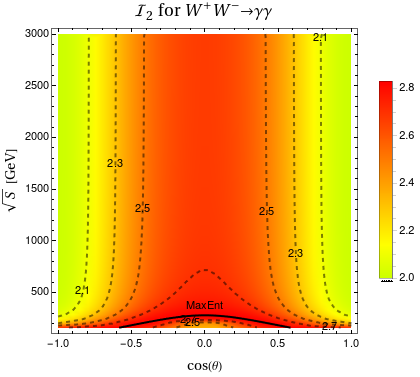}
  \end{tabular}
\caption{Negativity (left) and $\mI_2$ quantifier (right) for $\ww\to\gamma\gamma$ in the plane $[\cos(\theta),\sqrt{S}]$. Dashed contour lines are shown for an easy comparison of the numerical values. The solid contour line corresponds to the maximal Negativity equals to $1/2$ associated to Maximally Entangled states of \eqref{WWtoAA-maxent}.}
\label{WWtoAA-plot}
\end{figure}

The left panel of \figref{WWtoAA-plot} shows the behaviour of the Negativity for this process as a function of the energy and the scattering angle.
The minima of this quantifier is located at the upper corners, i.e., for $\cos(\theta)=\pm 1$ and large energies. From the analytical expression of \eqref{WWtoAA-Neg}, in this regime the Negativity behaves as $\sim 1.5\mw^2/S$ and reaches the minimum value $\sim 10^{-3}$ for $\sqrt{S}=3$ TeV. In particular, it never vanishes.
On the other hand, the arithmetic and geometric means inequality provides the maxima $1/2$ for the Negativity in \eqref{WWtoAA-Neg} when $U_1^{\gamma\gamma}=U_2^{\gamma\gamma}$. This maxima is the theoretical maximum expected for the Negativity corresponding to the Maximally Entangled pure states. The curve of Negativity equals to $1/2$ is given by
\be
S\vert_{\rm MaxEnt} = \frac{12(1-\cos^2(\theta))\mw^2}{1+3\cos^2(\theta)} 
\label{WWtoAA-maxent}
\ee
and it is represented by the solid black line in the red region of \figref{WWtoAA-plot}.

Furthermore, the Entropy of Entanglement of \eqref{definition-vNS} can also be computed analytically. The reduced matrices respect to each photon coincide and are given by  
\be
\rho_{red} = \begin{pmatrix}
\frac{1}{2} & \frac{8}{D_{\gamma\gamma}}(S-3\mw^2)((1+3c^2)S-12(1-c^2)\mw^2) \\
\frac{8}{D_{\gamma\gamma}}(S-3\mw^2)((1+3c^2)S-12(1-c^2)\mw^2) & \frac{1}{2}
\end{pmatrix}
\label{WWtoAA-redmatrix}
\ee
with eigenvalues equal to
\be
\lambda^{red}_1 = \frac{U^{\gamma\gamma}_1}{D_{\gamma\gamma}}  \quad{\rm and}\quad  \lambda^{red}_2 = \frac{U^{\gamma\gamma}_2}{D_{\gamma\gamma}}
\ee
Therefore, the resulting Entropy of Entanglement is
\be
\vNS = \frac{U^{\gamma\gamma}_1}{U_1^{\gamma\gamma}+U_2^{\gamma\gamma}}\log\left(1+\frac{U^{\gamma\gamma}_2}{U^{\gamma\gamma}_1}\right) + \frac{U^{\gamma\gamma}_2}{U_1^{\gamma\gamma}+U_2^{\gamma\gamma}}\log\left(1+\frac{U^{\gamma\gamma}_1}{U^{\gamma\gamma}_2} \right)
\label{WWtoAA-vNS}
\ee
The behaviour of this quantifier in the plane $[\cos(\theta),\sqrt{S}]$ is very similar to the Negativity and the resulting plot is relegated to the upper-left corner of \figref{SEEremainingplots} in \appref{App-moreplots} for saving space here. 
The minima are located for $\cos(\theta)=\pm 1$ and large energies decreasing as $\sim(2.25\mw^4/S^2)(1-\log(2.25\mw^4/S^2))$.
On the other hand, \eqref{WWtoAA-vNS} for $U^{\gamma\gamma}_1=U^{\gamma\gamma}_2$ reaches the maximum theoretical value $\log(2)$ corresponding to the Maximally Entangled state described by the curve in \eqref{WWtoAA-maxent} for which the reduced matrix is the half of the identity matrix. 

Finally, the decomposition of \eqref{rho-decom} corresponding to this density matrix have the non-vanishing coefficients
\bear
&& A_1 = B_1 =\frac{16}{D_{\gamma\gamma}}\left(S-3\mw^2\right)\left((1+3c^2)S-12(1-c^2)\mw^2\right)\,,  \nn\\
&& C_{11} = 1\,, \qquad C_{23} = C_{32} = -\frac{128\sqrt{2}}{D_{\gamma\gamma}}\left(c\sqrt{1-c^2}(S-3\mw^2)\mw\sqrt{S}\right)\,,  \nn\\
&& C_{22} = -C_{33} = \frac{2}{D_{\gamma\gamma}}\left((15-6c^2-9c^4)S^2-8(9+10c^2-7c^4)\mw^2S+144c^2(2-c^2)\mw^4\right)  
\label{WWtoAA-rhodecom}
\eear
Hence the two largest eigenvalues of $C^{\rm T}C$ are
\be
r_1 = 1  \quad{\rm and}\quad  
r_2 = 1-\frac{256}{D_{\gamma\gamma}^2}\left(S-3\mw^2\right)^2\left((1+3c^2)S-12(1-c^2)\mw^2\right)^2
\ee
resulting in the $\mI_2$ quantifier of \eqref{definition-I2} for this process as
\be
\mI_2 = 2\sqrt{2-\frac{256}{D_{\gamma\gamma}^2}\left(S-3\mw^2\right)^2\left((1+3c^2)S-12(1-c^2)\mw^2\right)^2}
\ee
The behaviour of this function in the plane $[\cos(\theta),\sqrt{S}]$ is shown in the right panel of \figref{WWtoAA-plot}. 
The minima correspond to large energies in the direction $\cos(\theta)=\pm 1$ and can be written as $\sim 2+9\mw^4/S^2$. In particular, this quantifier is always greater than 2 signaling a violation of Bell inequality in the whole kinematical plane for this observable.
Conversely, for the Maximally Entangled state described by the curve in \eqref{WWtoAA-maxent}, this quantifier reaches the theoretical maximum $2\sqrt{2}$ corresponding to the Cirelson bound~\cite{Cirelson:1980ry}. 
These remarks show that $\ww\to\gamma\gamma$ could be an ideal laboratory for a Bell inequality test among the VBS processes but it requires polarization measurements of the final photons\footnote{Although the polarization of high-energy photons is not currently measure in ATLAS and CMS, in contrast to the case of massive gauge bosons, the LHCb Collaboration performed analysis for photon polarization in $b$-baryon decays~\cite{LHCb:2021byf}. There are also proposals to study CP properties of the Higgs boson through the di-photon decay~\cite{Bishara:2013vya,Gritsan:2022php}.} (see for instance a related discussion in~\cite{Fabbrichesi:2022ovb}).

\begin{figure}[h]
  \centering
  \begin{tabular}{cc}
\includegraphics[width=0.42\textwidth]{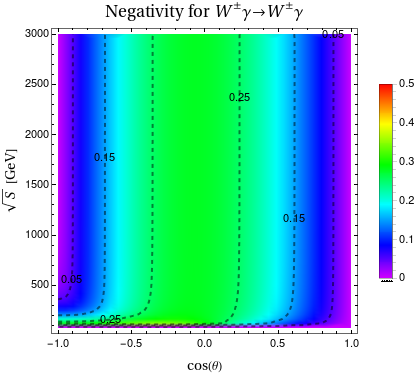} & \includegraphics[width=0.42\textwidth]{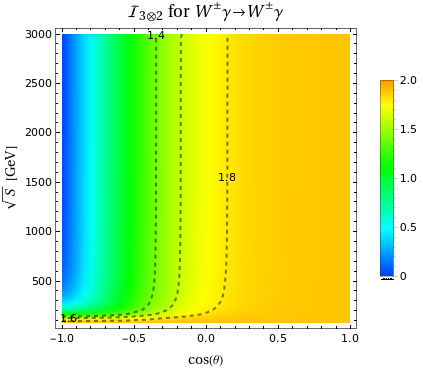} \\
\includegraphics[width=0.42\textwidth]{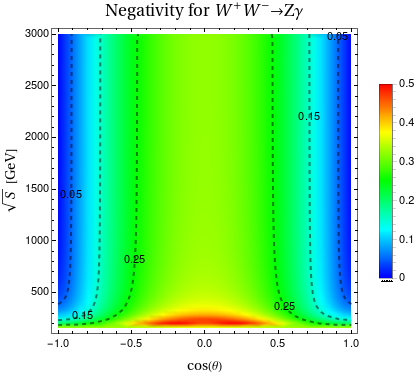} &  \includegraphics[width=0.42\textwidth]{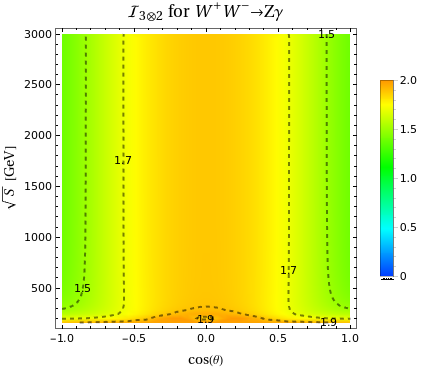}  \\
\includegraphics[width=0.42\textwidth]{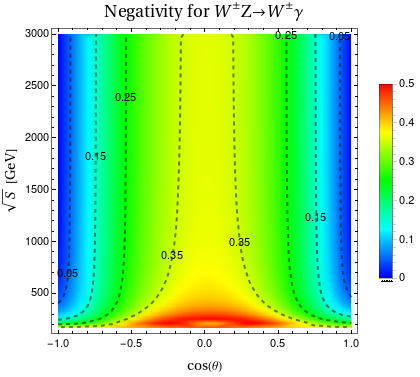} &  \includegraphics[width=0.42\textwidth]{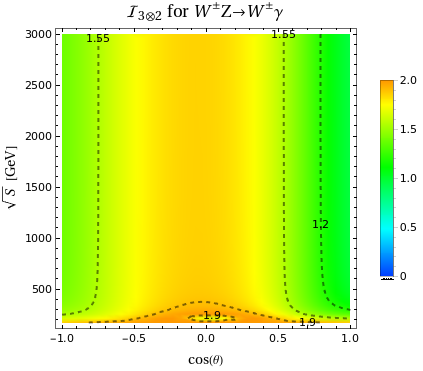}
  \end{tabular}
\caption{Negativity (left) and $\mI_{3\otimes 2}$ quantifier (right) for $\wpm\gamma\to\wpm\gamma$ (first row), $\ww\to Z\gamma$ (second row) and $\wpm Z\to\wpm\gamma$ (third row) in the plane $[\cos(\theta),\sqrt{S}]$. Contour lines are shown for an easy comparison of the numerical values.}
\label{qbqtplots}
\end{figure}

\subsection{qutrit$\otimes$qubit case}

The considered VBS processes of this kind are $\wpm\gamma\to\wpm\gamma$, $\ww\to Z\gamma$ and $\wpm Z\to\wpm\gamma$. The separability of these 3$\otimes$2 final states is also determined by the PPT criterion as in the previous section.
Just the $\wpm\gamma\to\wpm\gamma$ process was treated analytically and the coefficients of the decomposition in \eqref{rho-decom} are collected in \appref{App-analytic}. Remember the relevance of these coefficients due to their relation with the experimental data through quantum tomography. In addition, the analytical expression corresponding to the Entropy of Entanglement for this process is shown in this appendix.

The corresponding Negativity for $\wpm\gamma\to\wpm\gamma$ is presented in the left of the first row of \figref{qbqtplots}. In this process, the Negativity vanishes at the points $[\cos(\theta),\mw]$ and $[1,\sqrt{S}]$ of the plane $[\cos(\theta),\sqrt{S}]$.
For both kinematical regions, the six eigenvalues of the partial transpose density matrix are $\{1,0,0,0,0,0\}$. In consequence, the final state at the threshold and in the forward direction is separable. This fact can be also understood with the analytical expression of $\vNS$ in \eqref{WAtoWA-vNS}. In turn, the maximum value of the Negativity is $\sim$0.345 which indicates that the Maximally Entangled state never occurs in these scattering.
On the other hand, the Negativity for $\ww\to Z\gamma$ and $\wpm Z\to\wpm\gamma$ is shown in the left of the second and third rows of \figref{qbqtplots}.
Similar to the two-qubits case, the minima are located at the upper corners with values $\sim$8$\cdot 10^{-4}$. Also, the theoretical maximum $1/2$ is achieved for some points in the red region but the corresponding analytical curve cannot be computed.

The Entropy of Entanglement for these processes follows the same pattern of the Negativity, then corresponding plots are omitted here since no additional conclusions can be extracted. They are relegated to the \figref{SEEremainingplots} in \appref{App-moreplots} for saving space here. 

Regarding the Bell inequality, the quantifier $\mI_{3\otimes 2}$ of \eqref{definition-I3x2} is shown in the right column of \figref{qbqtplots} for each process.
This quantifier never exceeds 2 over the whole kinematical plane. Concretely, $\mI_{3\otimes 2}$ varies in the range $\sim(3\cdot10^{-3},1.88)$ for $\wpm\gamma\to\wpm\gamma$, in the range $\sim(1.38,2)$ for $\ww\to Z\gamma$ and $\sim(0.94,2)$ for $\wpm Z\to\wpm\gamma$. The maximum value 2 for the last two processes is reached for the points with maximal Nagativity.
Some comments are in order: firstly, the plots show that the entanglement is not a sufficient condition for Bell inequality violation. In particular, we have non-vanishing values of the Negativity but the corresponding points never exceeds 2 for the $\mI_{3\otimes 2}$ quantifier.
Secondly, these results are expected~\cite{Barr:2021zcp,Caban:2008qa} since the generalized CHSH operator of \eqref{eq-BellOP-genCHSH} is diminished by the vanishing outcome of the spin operator for massive gauge bosons.
As far as we know, there is no optimization for Bell operator in the $3\otimes 2$ case and it deserves a further study which also applied to single-top processes.

\subsection{two-qutrits case}

\begin{figure}[t]
    \centering
\begin{tabular}{cc}
\includegraphics[width=0.43\textwidth]{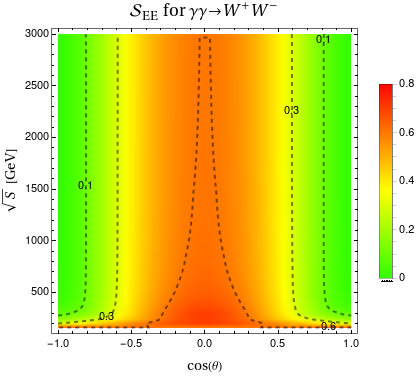} & \includegraphics[width=0.43\textwidth]{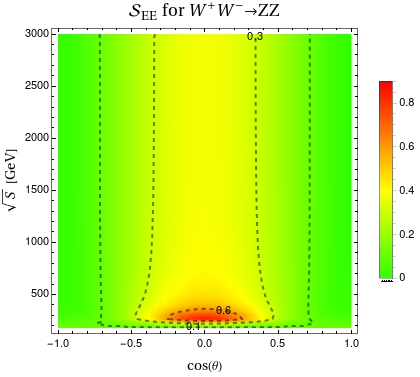}  \\
\includegraphics[width=0.43\textwidth]{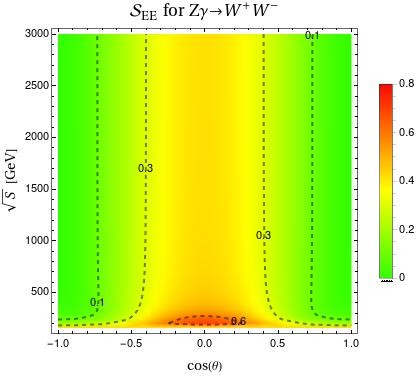} & \includegraphics[width=0.43\textwidth]{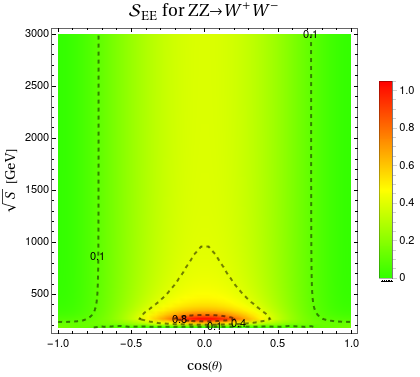}  \\
\end{tabular}
\includegraphics[width=0.43\textwidth]{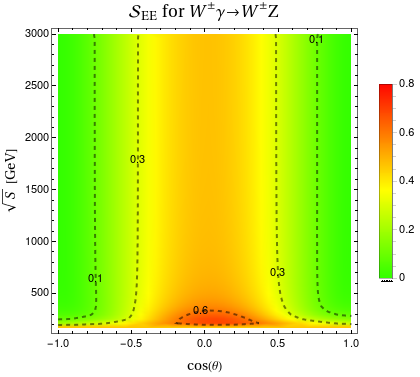}
\caption{Entropy of Entanglement for different VBS processes in the plane $[\cos(\theta),\sqrt{S}]$. Contour lines are shown for an easy comparison of the numerical values.}
\label{SEEplots1}
\end{figure}

The VBS processes corresponding to bipartite 3$\otimes$3 states have massive gauge bosons in the final state. Concretely, they are $\gamma\gamma\to\ww$, $\wpm\gamma\to\wpm Z$, $\ww\to\ww$, $\wpm\wpm\to\wpm\wpm$, $\ww\to ZZ$, $\wpm Z\to\wpm Z$, $Z\gamma\to\ww$, $ZZ\to\ww$ and $ZZ\to ZZ$. 
As presented in \secref{section-methods}, the level of entanglement of these pure states is given by non-vanishing values of the Entropy of Entanglement in \eqref{definition-vNS} and  Concurrence in \eqref{definition-Concu}.
The analytical expressions for the coefficients of the decomposition in \eqref{rho-decom} and the Entropy of Entanglement are given in \appref{App-analytic} just for $\gamma\gamma\to\ww$ process.

The \figref{SEEplots1} collects the $\vNS$ for $\gamma\gamma\to\ww$, $\ww\to ZZ$, $Z\gamma\to\ww$, $ZZ\to\ww$ and $\wpm\gamma\to\wpm Z$.
The green regions match with values lower than $\sim$0.1 and are located in both directions $\cos(\theta)\sim\pm 1$. As the energy increases, this entropy diminishes reaching values $\sim 10^{-5}$ but never vanishes. 
The red regions in the direction $\cos(\theta)\sim 0$ and near the threshold correspond to the maximal entropy values. In general, they are between 0.7 and 0.8 except for the $ZZ\to\ww$ (right panel of second row) having 1.04 which is closer to the theoretical maximum $\log(3)$ corresponding to the Maximally Entangled state.

\begin{figure}[t]
    \centering
\begin{tabular}{cc}
\includegraphics[width=0.44\textwidth]{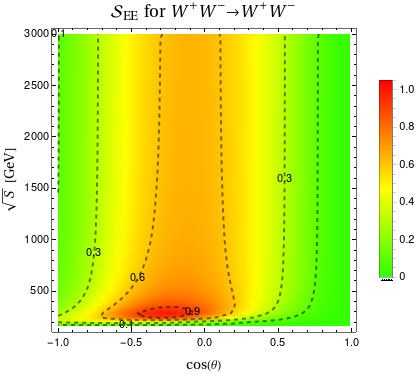} & \includegraphics[width=0.44\textwidth]{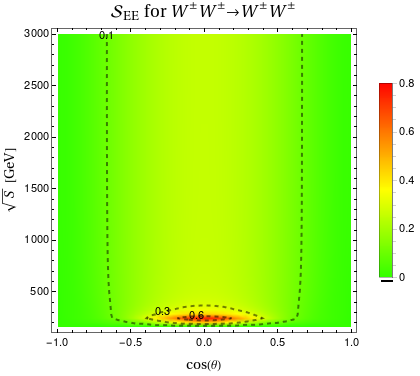}  \\
\includegraphics[width=0.44\textwidth]{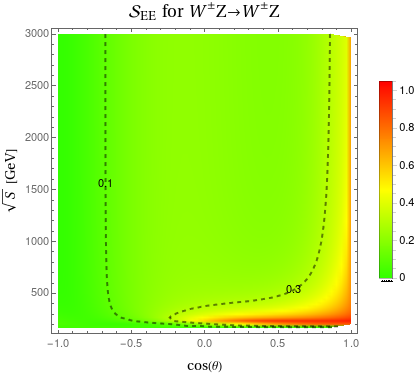} & \includegraphics[width=0.44\textwidth]{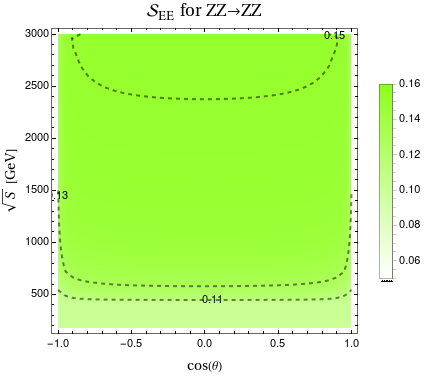}  \\
\end{tabular}
\caption{Entropy of Entanglement for the rest of VBS processes in the plane $[\cos(\theta),\sqrt{S}]$. Contour lines are shown for an easy comparison of the numerical values.}
\label{SEEplots2}
\end{figure}

The \figref{SEEplots2} gathers the rest of the VBS processes $\ww\to\ww$, $\wpm\wpm\to\wpm\wpm$, $\wpm Z\to\wpm Z$ and $ZZ\to ZZ$. As before, none of these processes have vanishing entropy.
The first two (showed in the first row) have the minima at the lower corners, i.e. near the threshold and not at large energies as for processes of \figref{SEEplots1}. 
The maxima are also near the threshold but in the direction $\cos(\theta)\sim 0$. 
The $\wpm Z\to\wpm Z$ (left panel of the second row) exhibits a strong asymmetry in the scattering angle, with lower values of the entropy for $\cos(\theta)\sim -1$ and minima located at large energies. On the contrary, the maxima is $\sim 1$ situated in the lower-right corner.
The later process has values between 0.1 and 0.15 in the whole plane in contrast to the large variations showed in the other VBS processes.

The behaviour of the Entanglement Entropy is also manifest in a very similar way for the Concurrence in each VBS process. The corresponding plots are relegated to \figrefs{Concuplots1}{Concuplots2} in \appref{App-moreplots}.
As before, this quantifier never reaches the zero value, i.e. the final states are entangled in the whole kinematical plane.
The maximal value of the Concurrence is 1.12 also for $ZZ\to\ww$ which is close to the theoretical maximum is $2/\sqrt{3}$.
\begin{figure}[t]
    \centering
\begin{tabular}{cc}
\includegraphics[width=0.45\textwidth]{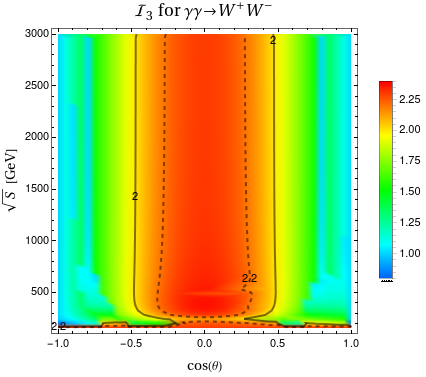} & \includegraphics[width=0.45\textwidth]{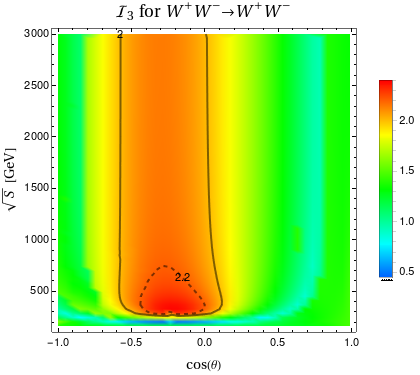}  \\
\end{tabular}
\includegraphics[width=0.45\textwidth]{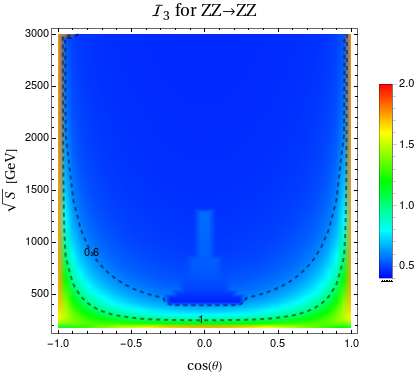} 
\caption{$\mI_3$ quantifier for $\gamma\gamma\to\ww$ (left of the first row), $\ww\to\ww$ (right of the first row) and $ZZ\to ZZ$ (second row) in the plane $[\cos(\theta),\sqrt{S}]$. The solid black contour line corresponds to $\mI_3$=2, limiting the region for Bell inequality violation. Dashed lines are shown for an easy comparison of the numerical values.}
\label{I3plots1}
\end{figure}

\begin{figure}[t]
    \centering
\begin{tabular}{cc}
\includegraphics[width=0.45\textwidth]{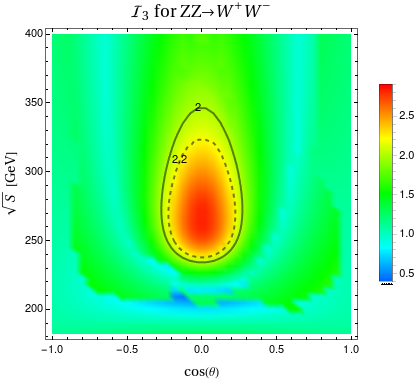} & \includegraphics[width=0.45\textwidth]{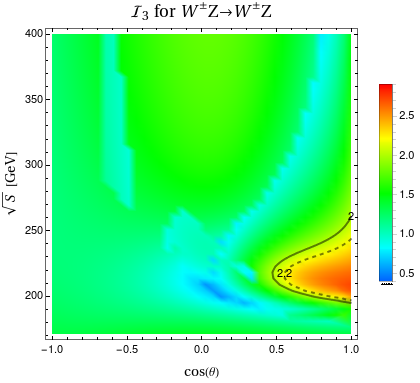}  \\
\includegraphics[width=0.45\textwidth]{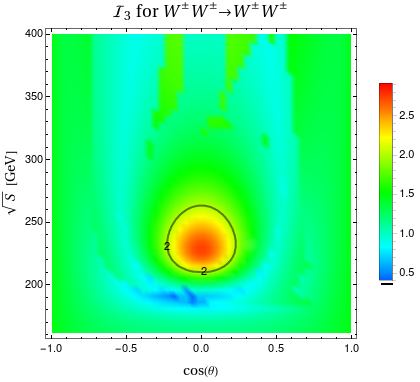} & \includegraphics[width=0.45\textwidth]{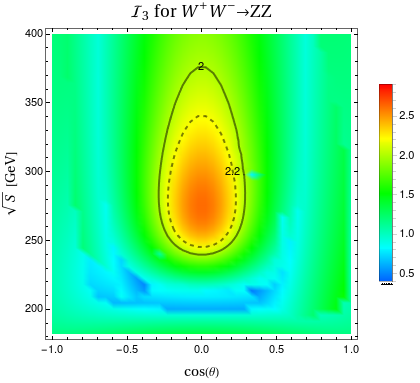}  \\
\includegraphics[width=0.45\textwidth]{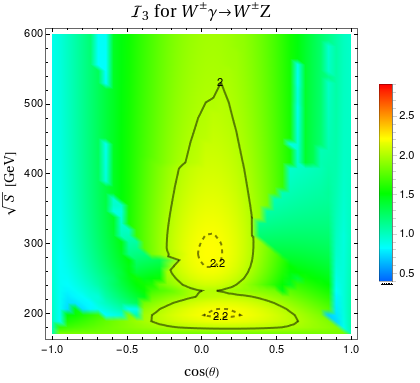} & \includegraphics[width=0.45\textwidth]{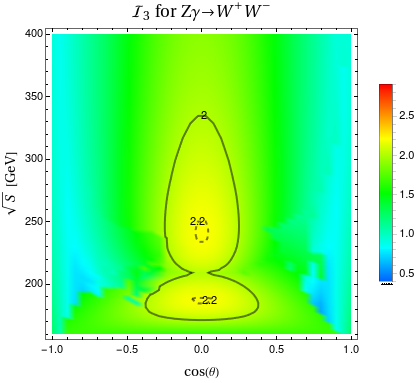} 
\end{tabular}
\caption{$\mI_3$ quantifier for the rest of VBS processes in the plane $[\cos(\theta),\sqrt{S}]$. The solid black contour line corresponds to $\mI_3$=2, limiting the region for Bell inequality violation. Dashed lines are shown for an easy comparison of the numerical values.}
\label{I3plots2}
\end{figure}

Finally, the theoretical predictions for the violation of Bell inequalities by means of the $\mI_3$ parameter in \eqref{definition-I3} are also presented in the plane $[\cos(\theta),\sqrt{S}]$ for each VBS process in this two-qutrits case. For each kinematical point, the rotation matrices $U_1$ and $U_2$ are determined in order to get maximal values for this quantifier.
The first row of \figref{I3plots1} shows the predictions for $\gamma\gamma\to\ww$ (left) and $\ww\to\ww$ (right). The solid black line corresponds to $\mI_3=2$ and delimits the orange-red region having values greater than 2. In both cases, the achieved maximum is $\sim$2.38 For energies above 300 GeV there is violation of Bell inequality when $-0.5\lesssim\cos(\theta)\lesssim 0.5$ ($\gamma\gamma\to\ww$) 
\vspace{5mm}

\hspace{-7.5mm}and $-0.6\lesssim\cos(\theta)\lesssim 0.05$ ($\ww\to\ww$).
On the contrary, the $ZZ\to ZZ$ (second row of \figref{I3plots1}) never exceeds 2 in the whole plane and reaches the maximum $\sim$1.76 in the upper corners.

For the rest of the VBS processes, the region corresponding to $\mI_3>2$ has energy below 600 GeV. Then they are collected in \figref{I3plots2} for this low energy region in which the solid black line also delimits the region for Bell inequality violation. 

The first row contains the $ZZ\to\ww$ and $\wpm Z\to\wpm Z$ having maximum $\sim$2.82 and $\sim$2.72, respectively. The maximum for $\wpm\wpm\to\wpm\wpm$ and $\ww\to ZZ$ (second row) are $\sim$2.71 and $\sim$2.62. In addition, $\wpm\gamma\to\wpm Z$ and $Z\gamma\to\ww$ are in the last row and both have maximum $\sim$2.21 
All these maxima are located in $\cos(\theta)\sim 0$ and energies between 230 GeV and 290 GeV depending on the process, except for $\wpm Z\to\wpm Z$ for which is located at $[\cos(\theta),\sqrt{S}]\sim$[1,210 GeV].

As discussed in the qutrit$\otimes$qubit processes, entanglement is just a necessary but not sufficient condition for Bell inequality violation. We showed that all the final states result entangled after the scattering process in the whole $[\cos(\theta),\sqrt{S}]$ plane but $\mI_3$ greater than 2 is just achieved in small regions of the phase space (even worse for $ZZ\to ZZ$ which never violates the Bell inequality).
\vspace{5mm}

\begin{figure}[h]
    \centering
\includegraphics[width=0.45\textwidth]{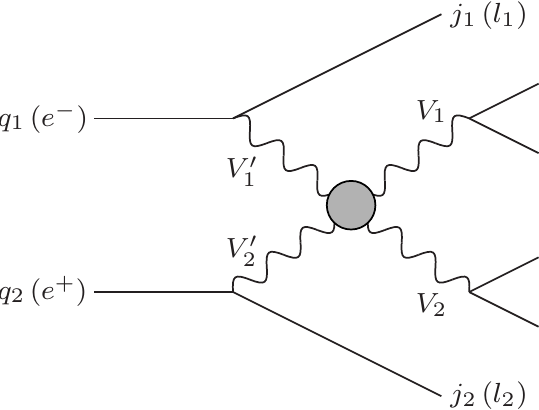}
\caption{Experimental signature for VBS at sub-process level in the LHC (future electron-positron colliders). In the latter case, the companion leptons $l_{1,2}$ can be electrons, positrons or neutrinos depending on the gauge bosons $V'_{1,2}$. The leptonic decays of the $V_{1,2}$ are represented by the final solid lines.}
\label{fig-expsignature}
\end{figure}

\subsection{Prospects at colliders}

The analyzed VBS channels correspond to a sub-process at collider, yielding to the following experimental signatures:
\bear
pp &\to& (V_1V_2)_{\rm leptonic\,decay} +j_1+j_2 \quad{\rm for\,\,the\,\,LHC,}  \nn\\
e^-e^+ &\to& (V_1V_2)_{\rm leptonic\,decay} +l_1+l_2 \quad{\rm for\,\,future\,\,electron-positron\,\,colliders,}
\label{expsignature}
\eear
which are generically represented in \figref{fig-expsignature}. 

In this figure, the gray circle corresponds to the tree level VBS processes described in \appref{App-kinem}.
This kind of processes are purely EW and very rare at the LHC, however they allow to probe the core of the EW Symmetry Breaking mechanism due to the interplay among the triple and quartic gauge self-coupling with the Higgs boson interactions. Nowadays, measurements of fiducial and total cross-sections as well as polarization of the outgoing gauge bosons are performed by ATLAS and CMS Collaborations~\cite{Covarelli:2021fra}.
The reported results with the 13 TeV data for all\footnote{$\ww\to\gamma\gamma$ process has not been measured yet at the LHC, due to the immense QCD multijet background~\cite{Covarelli:2021gyz}. However recent sensibility studies can be found in~\cite{Forster:2842604}.} the considered final vector boson pairs in this work, in association with two jets, are: $\wpm\gamma$~\cite{CMS:2022yrl}, $Z\gamma$~\cite{CMS:2021gme,ATLAS:2023fxh}, $\wpm\wpm$~\cite{ATLAS:2019cbr,CMS:2020etf}, $\wpm Z$~\cite{ATLAS:2018mxa,ATLAS:2019thr,CMS:2020gfh,CMS:2021qzz}, $\ww$~\cite{ATLAS:2020iwi,CMS:2022woe} and $ZZ$~\cite{ATLAS:2020nlt,CMS:2020fqz}.

The LHC signal has a very characteristic kinematics at detector level given by two energetic jets in the forward region with large invariant mass $m_{j_1j_2}$ and a wide pseudo-rapidity separation $\Delta\eta_{j_1j_2}$. 
A similar topology for the companion leptons in the future electron-positron colliders is expected~\cite{BuarqueFranzosi:2021wrv}.
The relevant decay modes for the final gauge bosons are the leptonic ones since the momenta of these leptons provide a measurement of the gauge boson polarizations, see for instance~\cite{Rahaman:2021fcz} and also the references therein. 
In particular, the corresponding angular distributions of the decays lead to a reconstruction of the density matrix from the experimental data as it was developed in~\cite{Ashby-Pickering:2022umy}.

The aim of this work was to perform a theoretical determination of the density matrix and the corresponding entanglement quantifiers for different VBS processes, which are not explored in the literature as far as we know. 
Notice that at the complete process level in \eqref{expsignature}, the resulting density matrix represents a mixed-state at collider~\cite{Afik:2022kwm} via a convex combination of the density matrices for each unpolarized gauge boson $V'_{1,2}$ cases considering them as `partons' inside the initial fermions. 
In particular, this mixed density matrix could be computed by means of the Weizsacker-Williams Approximation~\cite{vonWeizsacker:1934nji,PhysRev.45.729} for photons and the Effective $W$ Approximation~\cite{Dawson:1984gx,PhysRevD.36.291} for massive vector bosons. 
In that case just a lower bound of the Concurrence can be determined~\cite{Fabbrichesi:2023cev} in order to decide if the system $V_1\otimes V_2$ is entangled or separable.
In addition, the experimental challenges in each production mechanism and for each collider are different then, dedicated studies are mandatory.
This computation and the corresponding reconstruction of the entanglement quantifiers from Monte-Carlo simulation is beyond the scope of this work and it is postponed for a future study.

\section{Summary and perspectives}
\label{section-conclus}

In this work, the quantum properties of vector boson scattering were explored by the computation of the quantifiers associated to entanglement/separability and violation of Bell inequalities.
In particular, the Negativity $\mN$, Entropy of Entanglement $\vNS$ and Concurrence $\mC$ can be treated as entanglement/separability parameters, in a sense that they characterize a degree of entanglement, whereas the violation of Bell inequalities is determined by the $\mI$ parameter corresponding to CHSH or CGLMP operators.
This kind of processes allows to examine pure bipartite systems, associated to the spin of the final vector bosons, with density matrix of dimensions 2$\otimes$2 (for $\ww\to\gamma\gamma$), 3$\otimes$2 (for $\wpm\gamma\to\wpm\gamma$, $\ww\to Z\gamma$ and $\wpm Z\to\wpm\gamma$) and 3$\otimes$3 (for $\gamma\gamma\to\ww$, $\wpm\gamma\to\wpm Z$, $\ww\to\ww$, $\wpm\wpm\to\wpm\wpm$, $\ww\to ZZ$, $\wpm Z\to\wpm Z$, $Z\gamma\to\ww$, $ZZ\to\ww$ and $ZZ\to ZZ$). 
The corresponding amplitudes were computed at tree level in the context of the SM. Analytical expressions of density matrix and entanglement quantifiers were presented for $\ww\to\gamma\gamma$, $\wpm\gamma\to\wpm\gamma$ and $\gamma\gamma\to\ww$.

The goal of this paper was to determine the kinematical region in the plane $[\cos(\theta),\sqrt{S}]$ where the resulting final vector bosons after the scattering were entangled and then, if is it possible to test the Bell inequality in that region.
We found that all the final states are entangled after the scattering, except for $\wpm\gamma\to\wpm\gamma$ in the forward direction $\cos(\theta)=1$ or at energy equal to $\sqrt{S}=\mw$. 
Also, for $\ww\to\gamma\gamma$, $\ww\to Z\gamma$ and $\wpm Z\to\wpm\gamma$ processes, the Maximally Entangled state is reached in particular kinematical configuration since the entanglement quantifiers achieve the maximum theoretical values there.
On the other hand, compared to the rest of the VBS processes, the $ZZ\to ZZ$ has the less entangled final state in the whole kinematical plane.

Regarding the Bell inequality, we conclude that in the whole range of the scattering angle and up to energy of 3 TeV, the CHSH $\mI_2$ parameter for $\ww\to\gamma\gamma$ is greater than 2 and the Maximally Entangled state of \eqref{WWtoAA-maxent} provides the maximal violation, corresponding to the Cirelson bound 2$\sqrt{2}$.  
In addition, Bell inequality violation is expected for the rest of the VBS processes but it occurs in small regions of the phase space, except for the qutrit$\otimes$qubit processes nor $ZZ\to ZZ$ which the corresponding Bell parameters are always lower than 2.
Therefore, it is manifest that entanglement is not a sufficient condition for Bell inequality violation.
It is important to stress that the implemented $\mI_{3\otimes2}$ quantifier was not optimized for these VBS processes in the sense that the corresponding Bell operator is the generalization of the CHSH to higher dimension. Analogously, $\mI_3$ was optimized for Higgs boson decays into massive gauge bosons. Therefore, the determination of more appropriate Bell operators for qutrit$\otimes$qubit and two-qutrits VBS processes is an interesting improvement for future works.

The previous theoretical predictions intend to guide the experimental search for the quantum properties of VBS in the kinematical plane since different reconstruction techniques must be implemented depending on how boosted are the final particles or if the available energy is near the threshold or much higher.
This study is a first step in that direction for this kind of processes.
At collider level, they correspond to sub-process in a more complex scattering yielding to a mixed system for which the computation of the entanglement quantifiers results in a non-trivial maximization problem over a convex sum of the analyzed bipartite systems.
The next step is to perform this computation using the Effective W Approximation in order to get predictions for the LHC and future lepton colliders.
In addition, a quantum tomography analysis with Monte-Carlo simulations will be developed to estimate the significances to these observables, as it was done for Higgs boson decay and diboson production in~\cite{Aguilar-Saavedra:2022wam,Ashby-Pickering:2022umy,Fabbrichesi:2023cev}.

Other lines for further studies are related to BSM physics. Similar to EFT analysis in~\cite{Aoude:2022imd,Fabbrichesi:2022ovb,Severi:2022qjy,Fabbrichesi:2023jep} or demanding maximal entanglement as fundamental principle~\cite{Cervera-Lierta:2017tdt}, quantum information measurements on VBS processes could constrain new physics operators in the EWSB sector not explored yet.

\section*{Acknowledgments}

I am grateful to Bianca Polari who encouraged the abidance of this study. I also thank to Alejandro Szynkman, Ernesto Arganda and Francisco Alonso for their helpful discussions and suggestions.
The present work has received financial support from CONICET and ANPCyT under projects PICT 2017-2751, PICT 2018-03682 and PICT-2021-I-INVI-00374.
\newpage

\section*{Appendices}
\appendix

\section{VBS kinematics and amplitudes}
\label{App-kinem}

This appendix is devoted to summarize the relevant details for the computation of the amplitudes corresponding to VBS processes in \eqref{eq-VBSproc}. 
Without loss of generality, the center-of-mass (CM) frame is chosen with the incoming particles traveling along the $z$ axis and then are scattered into the $x-z$ plane with angle $\theta$. In that case, the 2$\to$2 amplitudes can be written in terms of two kinematical variables: the CM energy $\sqrt{S}$ and the scattering angle $\cos(\theta)$. 
Explicitly, the momenta $p'_i$, $p_i$ and the polarization vectors $\varepsilon'_i(s'_i)$, $\varepsilon_i(s_i)$ with the usual normalizations are
\bear
p'_1 &=& \left(E'_1,0,0,-p_{in}\right) \hspace{20mm} p_1 = \left(E_1,-p_{out}\sin(\theta),0,-p_{out}\cos(\theta)\right)  \nn\\
\varepsilon'_1(\pm) &=& \frac{1}{\sqrt2}\left(0,-i,\mp1,0\right) \hspace{12mm} \varepsilon^*_1(\pm) = \frac{1}{\sqrt2}\left(0,i\cos(\theta),\mp1,-i\sin(\theta)\right)  \nn\\
\varepsilon'_1(0) &=& \frac{1}{m_{V'_1}}\left(p_{in},0,0,-E'_1\right) \hspace{8mm} \varepsilon^*_1(0) = \frac{1}{m_{V_1}}\left(p_{out},-E_1\sin(\theta),0,-E_1\cos(\theta)\right)  \nn\\
&&\nn\\
p'_2 &=& \left(E'_2,0,0,p_{in}\right) \hspace{23mm} p_2 = \left(E_2,p_{out}\sin(\theta),0,p_{out}\cos(\theta)\right)  \nn\\
\varepsilon'_2(\pm) &=& \frac{1}{\sqrt2}\left(0,-i,\mp1,0\right) \hspace{12mm} \varepsilon^*_2(\pm) = \frac{1}{\sqrt2}\left(0,i\cos(\theta),\mp1,-i\sin(\theta)\right)  \nn\\
\varepsilon'_2(0) &=& \frac{1}{m_{V'_2}}\left(-p_{in},0,0,-E'_2\right) \hspace{6mm} \varepsilon^*_2(0) = \frac{1}{m_{V_2}}\left(-p_{out},-E_2\sin(\theta),0,-E_2\cos(\theta)\right)
\eear
with the trimomentum and energies given by
\bear
p_{in} &=& \frac{1}{2}\sqrt{\lambda(S,m_{V'_1}^2,m_{V'_2}^2)/S}\,,  \hspace{8mm} E'_i=\sqrt{m_{V'_i}^2+p_{in}^2}\,,  \nn\\
p_{out} &=& \frac{1}{2}\sqrt{\lambda(S,m_{V_1}^2,m_{V_2}^2)/S}\,,  \hspace{8mm} E_i=\sqrt{m_{V_i}^2+p_{out}^2}
\eear
where the Kallen function is $\lambda(a,b,c)=a^2+b^2+c^2-2ab-2bc-2ca$.
The polarizations of the gauge bosons define the conventional basis, corresponding to the third component of the spin, as $\{\vert+\rangle,\vert-\rangle\}$ for photons and $\{\vert+\rangle,\vert0\rangle,\vert-\rangle\}$ for $\wpm$ and $Z$ bosons.

\begin{figure}
    \centering
\includegraphics[width=0.9\textwidth]{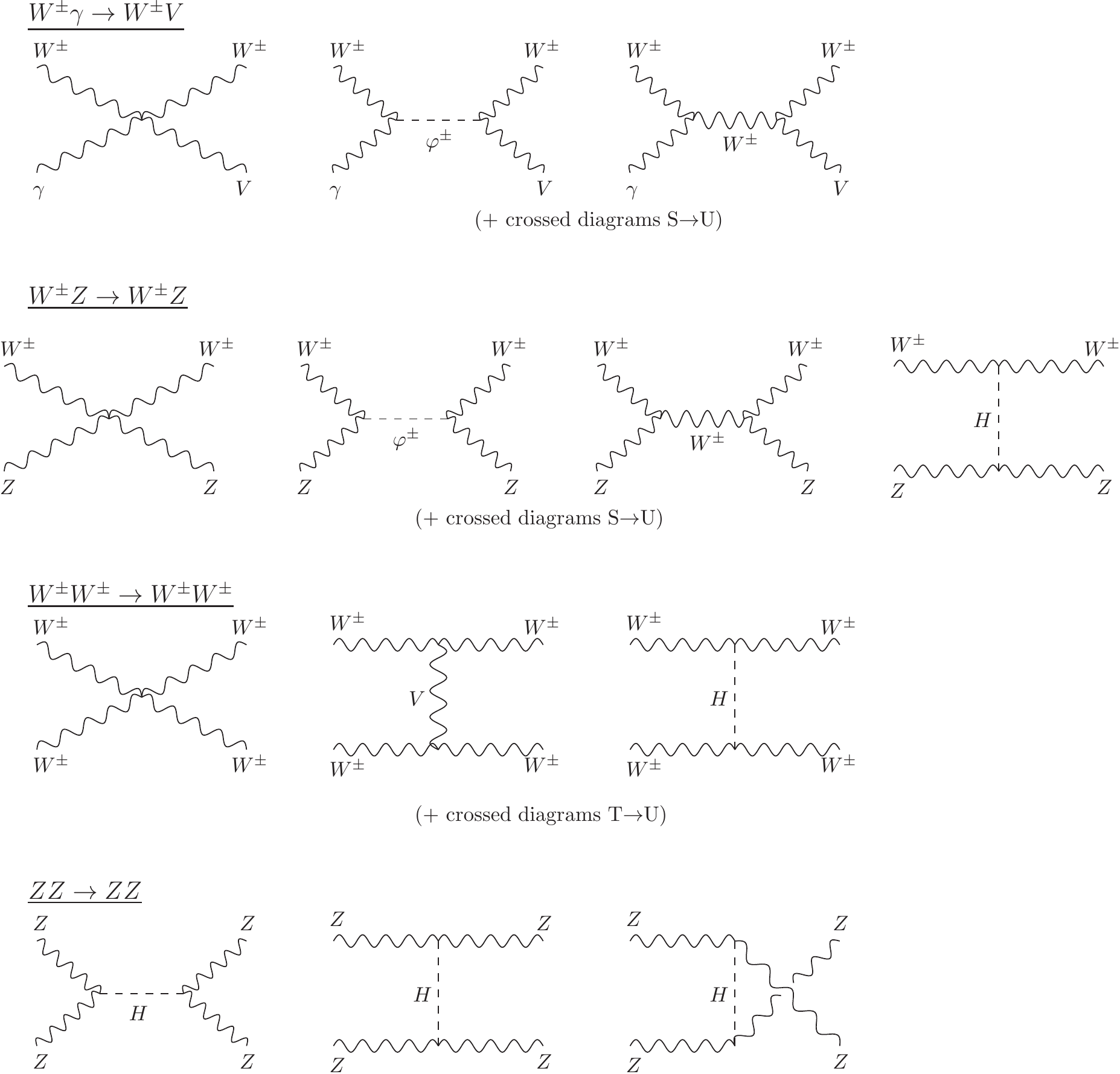}
\caption{Representative Feynman diagrams for some of the considered VBS channels ($V$ stands for photon and $Z$ boson). Processes related by crossing-symmetry are described in the text.}
\label{feyndiags}
\end{figure}

The SM amplitudes were computed in the Feynman-'t Hooft gauge at tree level. The $\wpm\gamma\to\wpm\gamma$ and $\wpm\gamma\to\wpm Z$ have the same Feynman diagrams, which are presented in the first row of \figref{feyndiags} as $\wpm\gamma\to\wpm V$. Diagrams with $\wpm$ and charged Goldstone $\varphi^\pm$ bosons in the $U$-channel are omitted for simplicity. Notice that the Higgs boson is not present in these processes. Of course, $\ww\to\gamma V$ and $\wpm Z\to\wpm\gamma$ are related by crossing symmetry and have similar diagrams. 
For $\wpm Z\to\wpm Z$, and the related processes by crossing symmetry $\ww\to ZZ$ and $ZZ\to\ww$, there is an additional diagram with the Higgs boson as mediator respect to the previous ones as can be seen in the second row.
On the other hand, the $\ww\to\ww$ and $\wpm\wpm\to\wpm\wpm$ processes in the third row of \figref{feyndiags} have two diagrams with the Higgs boson as mediator.
Finally, the $ZZ\to ZZ$ process only has 3 diagrams with the Higgs boson as mediator in $S$-, $T$- and $U$-channels, which are shown in the last row. 

\section{Generalized Gell-Mann and spin matrices for dimension 2 and 3}
\label{App-matrices}

For completeness, this appendix gathers the explicit form of the generalized Gell-Mann matrices that enter in the decomposition of the density matrix in \eqref{rho-decom}.
An important relation is the trace orthogonality of these matrices:
\be
{\rm Tr}\big[\lambda^{(d)}_i\big]=0 \quad\text{and}\quad {\rm Tr}\big[\lambda^{(d)}_i\lambda^{(d)}_j\big]=2\delta_{ij}
\label{eq-TrgGM}
\ee

For qubits, the dimension-2 Pauli matrices are
\be
\sigma_1 = \lambda_1^{(2)} = \begin{pmatrix}
0 & 1 \\
1 & 0 
\end{pmatrix}\,,  \hspace{8mm}  \sigma_2 = \lambda_2^{(2)} = \begin{pmatrix}
0 & -i \\
i & 0 
\end{pmatrix}\,,  \hspace{8mm}  \sigma_3 = \lambda_3^{(2)} = \begin{pmatrix}
1 & 0 \\
0 & -1 
\end{pmatrix}
\ee
and also the dimension-3 representation of the eight Gell-Mann matrices (for simplicity, the superscript `(3)' is omitted) are
\bear
\lambda_1 &=& \begin{pmatrix}
0 & 1 & 0\\
1 & 0 & 0\\
0 & 0 & 0
\end{pmatrix}\,,  \hspace{5mm}  \lambda_2 = \begin{pmatrix}
0 & -i & 0\\
i & 0 & 0\\
0 & 0 & 0
\end{pmatrix}\,,  \hspace{5mm}  \lambda_3 = \begin{pmatrix}
1 & 0 & 0\\
0 & -1 & 0\\
0 & 0 & 0
\end{pmatrix}\,,  \hspace{5mm}  \lambda_4 = \begin{pmatrix}
0 & 0 & 1\\
0 & 0 & 0\\
1 & 0 & 0
\end{pmatrix}\,,  \nn\\  
\lambda_5 &=& \begin{pmatrix}
0 & 0 & -i\\
0 & 0 & 0\\
i & 0 & 0
\end{pmatrix}\,,  \hspace{5mm}  \lambda_6 = \begin{pmatrix}
0 & 0 & 0\\
0 & 0 & 1\\
0 & 1 & 0
\end{pmatrix}\,,  \hspace{5mm}  \lambda_7 = \begin{pmatrix}
0 & 0 & 0\\
0 & 0 & -i\\
0 & i & 0
\end{pmatrix}\,,  \hspace{5mm}  \lambda_8 = \frac{1}{\sqrt{3}}\begin{pmatrix}
1 & 0 & 0\\
0 & 1 & 0\\
0 & 0 & -2
\end{pmatrix}
\eear

In addition, the spin-1 matrices in \eqref{eq-CGLMPxy} are
\bear
&&S_x = \frac{1}{\sqrt{2}}(\lambda_1+\lambda_6) = \frac{1}{\sqrt{2}}\begin{pmatrix}
0 & 1 & 0\\
1 & 0 & 1\\
0 & 1 & 0
\end{pmatrix}\,,  \hspace{7mm}  S_y = \frac{1}{\sqrt{2}}(\lambda_2+\lambda_7) = \frac{1}{\sqrt{2}}\begin{pmatrix}
0 & -i & 0\\
i & 0 & -i\\
0 & i & 0
\end{pmatrix}\,,  \nn\\
&&\hspace{30mm}  S_z = \frac{1}{2}(\lambda_3+\sqrt{3}\lambda_8) = \begin{pmatrix}
1 & 0 & 0\\
0 & 0 & 0\\
0 & 0 & -1
\end{pmatrix}
\label{eq-spinmatrices}
\eear

\section{Additional plots of entanglement quantifiers}
\label{App-moreplots}

\begin{figure}[t]
  \centering
  \begin{tabular}{cc}
\includegraphics[width=0.42\textwidth]{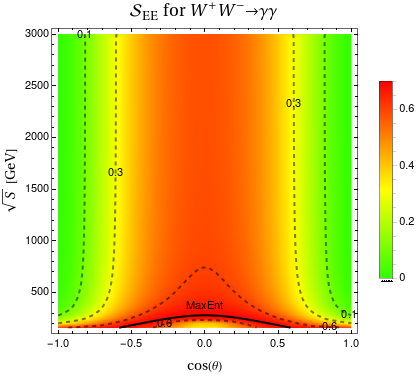} & \includegraphics[width=0.42\textwidth]{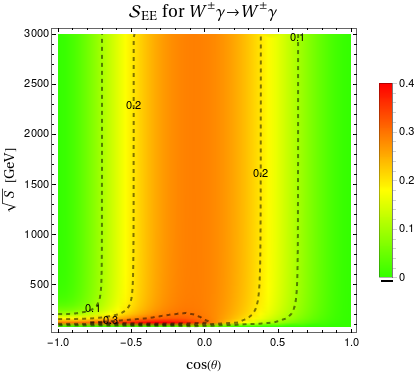} \\
\includegraphics[width=0.42\textwidth]{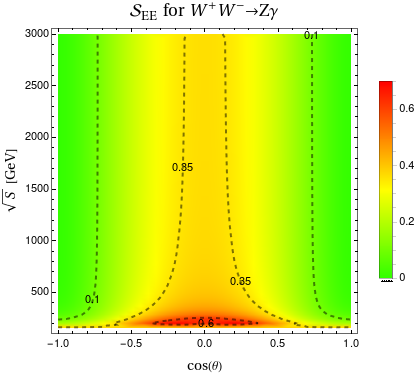} &  \includegraphics[width=0.42\textwidth]{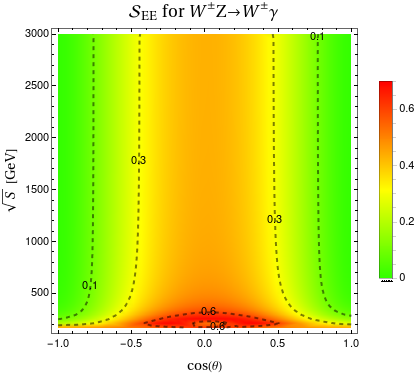}  
  \end{tabular}
\caption{Entropy of Entanglement for $\ww\to\gamma\gamma$ (upper-left), $\wpm\gamma\to\wpm\gamma$ (upper-right), $\ww\to Z\gamma$ (lower-left) and $\wpm Z\to\wpm\gamma$ (lower-right) in the plane $[\cos(\theta),\sqrt{S}]$. Contour lines are shown for an easy comparison of the numerical values.}
\label{SEEremainingplots}
\end{figure}

This appendix collects the plots of the quantifiers that were omitted in the main text for saving space. \figref{SEEremainingplots} corresponds to the Entropy of Entanglement for the two-qubits and qutrit$\otimes$qubit processes. As for the Negativity of these VBS, $\vNS$ never vanishes in the kinematical plane then the final states are entangled\footnote{Except for $\wpm\gamma\to\wpm\gamma$ at the threshold and in the forward direction as can be seen from \eqref{WAtoWA-vNS}.}. The maximum theoretical values for this quantifier, corresponding to the Maximally Entangled states, are achieved in $\ww\to\gamma\gamma$ (showed in solid black line in the upper-left plot), and in both $\ww\to Z\gamma$ and $\wpm Z\to\wpm\gamma$ (second row of this figure). 

\begin{figure}[t]
    \centering
\begin{tabular}{cc}
\includegraphics[width=0.45\textwidth]{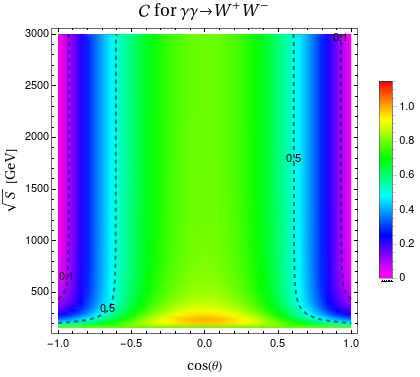} & \includegraphics[width=0.45\textwidth]{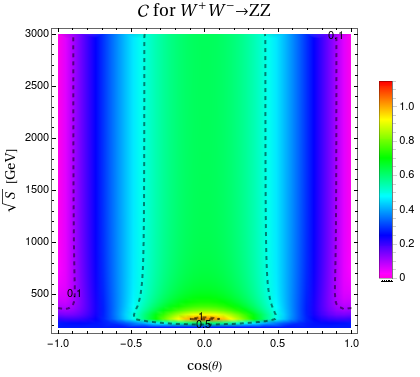}  \\
\includegraphics[width=0.45\textwidth]{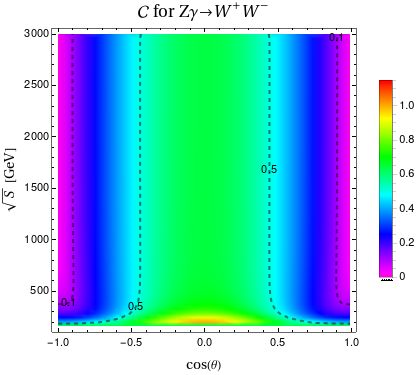} & \includegraphics[width=0.45\textwidth]{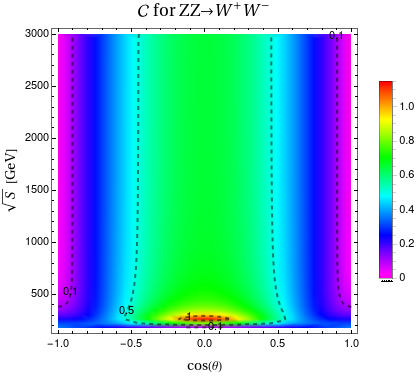}  \\
\end{tabular}
\includegraphics[width=0.45\textwidth]{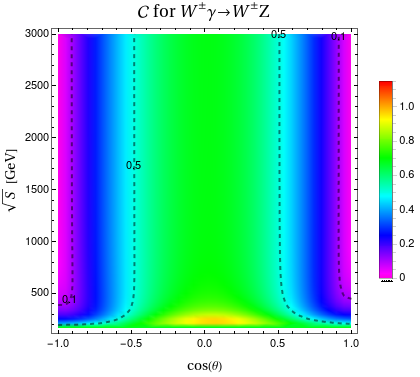}
\caption{Concurrence for different 3$\otimes$3 VBS processes in the plane $[\cos(\theta),\sqrt{S}]$. Contour lines are shown for an easy comparison of the numerical values.}
\label{Concuplots1}
\end{figure}

\figrefs{Concuplots1}{Concuplots2} show the Concurrence for all the processes with massive gauge bosons in final state, i.e. bipartite 3$\otimes$3 system. The same scale for all the plots is used, making more transparent the comparison among them.
Now, the lower values are denoted in violet colour and never reach the zero value, i.e. the final states are entangled in the whole kinematical plane.
The maximal value of the Concurrence is 1.12 for $ZZ\to\ww$ which is close to the theoretical maximum is $2/\sqrt{3}$.
\textcolor{white}{need space \vspace{20mm}}
\begin{figure}[t]
    \centering
\begin{tabular}{cc}
\includegraphics[width=0.45\textwidth]{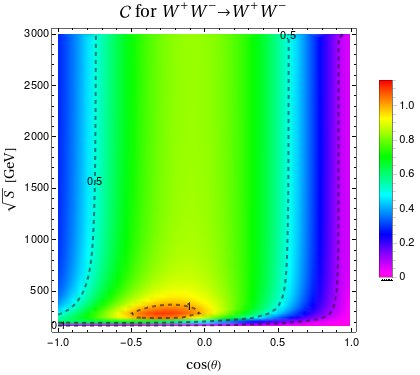} & \includegraphics[width=0.45\textwidth]{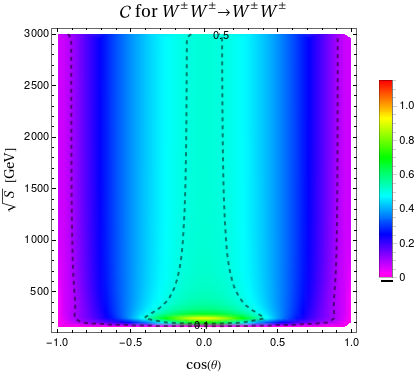}  \\
\includegraphics[width=0.45\textwidth]{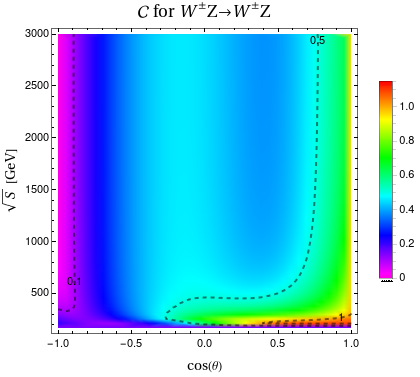} & \includegraphics[width=0.45\textwidth]{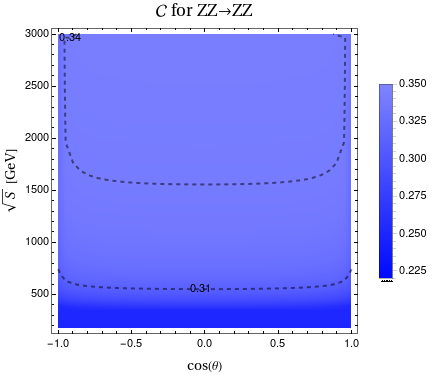}   \\
\end{tabular}
\caption{Concurrence for the rest of 3$\otimes$3 VBS processes in the plane $[\cos(\theta),\sqrt{S}]$. Contour lines are shown for an easy comparison of the numerical values.}
\label{Concuplots2}
\end{figure}

\textcolor{white}{need space \vspace{20mm}}

\section{Analytical expressions for $\wpm\gamma\to\wpm\gamma$ and $\gamma\gamma\to\ww$}
\label{App-analytic}

In this appendix, the analytical expressions of the non-vanishing coefficients in the decomposition of \eqref{rho-decom} for the density matrix corresponding to $\wpm\gamma\to\wpm\gamma$ and $\gamma\gamma\to\ww$ are presented. In addition, the eigenvalues of the reduced matrices that enter in the computation of the Entanglement Entropy in \eqref{definition-vNS} are also shown.
These observables are written in terms of the kinematical variables $\sqrt{S}$ and $c=\cos(\theta)$ and allow us to understand the numerical results.

\subsection{$\wpm\gamma\to\wpm\gamma$ process}

For this qutrit$\otimes$qubit case, the decomposition of the 6$\times$6 density matrix is
\be
\rho = \frac{1}{6}I_{6} +\frac{1}{4}\sum_{i=1}^{8}A_i\lambda_i\otimes I_{2} +\frac{1}{6}\sum_{j=1}^{3}B_j I_{3}\otimes\sigma_j +\frac{1}{4}\sum_{i=1}^{8}\sum_{j=1}^{3}C_{ij} \lambda_i\otimes\sigma_j 
\ee

For a compact notation, we define the quantity
\bear
D_{W\gamma} &=& (1-c)^4 \mw^8 -2 (1-c)^4 \mw^6 S +(1-c)^2 \left(2 c^2+7\right) \mw^4 S^2   \nn\\
&&-2 \left(c^4-3 c^2-4 c+6\right) \mw^2 S^3+\left(c^4+3 c^2+2 c+6\right) S^4
\eear

The resulting non-vanishing coefficients $A_i$ and $B_j$ are
{\allowdisplaybreaks
\bear
A_1 = A_6 &=& \frac{1}{D_{W\gamma}}S\left( (1-c)^3 \mw^6+(1-c)^2 (3 c+1) \mw^4 S \right.  \nn\\
&&\left.\hspace{8mm}-\left(3 c^3-9 c^2+c+5\right) \mw^2 S^2+\left(c^3+c^2+3 c+3\right) S^3 \right)  \nn\\
A_2 = -A_7 &=& \frac{1}{D_{W\gamma}}\sqrt{2(1-c^2)S}\mw\left( -(1-c)^3 \mw^6+(1-c)^2 (c+4) \mw^4 S \right.  \nn\\
&&\left.\hspace{8mm}+\left(c^3+c^2+3 c-5\right) \mw^2 S^2 +\left(c^3+c+2\right) S^3 \right)  \nn\\
A_3 = -\sqrt{3}A_8 &=& \frac{1}{2D_{W\gamma}}\left( (1-c)^4 \mw^8-4 (1-c)^3 (c+2) \mw^6 S \right.  \nn\\
&&\left.\hspace{10mm}+2 (1-c)^2 \left(c^2+6 c+8\right) \mw^4 S^2 +4 \left(c^4+2 c-3\right) \mw^2 S^3+\left(c^4-4 c+3\right) S^4 \right)  \nn\\
A_4 &=& \frac{2}{D_{W\gamma}}S\left( -(1-c)^3 (c+1) \mw^6 +2 \left(c^3-3 c+2\right) \mw^4 S \right.  \nn\\
&&\left.\hspace{8mm}+\left(c^4+4 c-5\right) \mw^2 S^2 +2 \left(c^2+1\right) S^3 \right)  \nn\\
B_1 &=& \frac{1}{D_{W\gamma}}S^2\left( 5 (1-c)^2 \mw^4 +2 \left(c^2+4 c-5\right) \mw^2 S+\left(5 c^2+2 c+5\right) S^2 \right) 
\eear
}
and the non-vanishing coefficients $C_{ij}$ of the correlation matrix are
{\allowdisplaybreaks
\bear
C_{11} = C_{61} &=& \frac{1}{D_{W\gamma}}S\left( (1-c)^3 \mw^6+(1-c)^2 (3 c+1) \mw^4 S \right.  \nn\\
&&\left.\hspace{8mm}-\left(3 c^3-9 c^2+c+5\right) \mw^2 S^2+\left(c^3+c^2+3 c+3\right) S^3  \right)  \nn\\
C_{12} = C_{23} = -C_{62} = C_{73} &=& -\frac{1}{D_{W\gamma}}(1-c)\sqrt{2(1-c^2)S}\mw(S-\mw^2)\left( (1-c)^2 \mw^4 \right.  \nn\\
&&\left.\hspace{12mm}-(1-c) \mw^2 S+c (c+1) S^2  \right)  \nn\\
C_{13} = C_{22} = -C_{63} = -C_{72} &=& \frac{1}{D_{W\gamma}}(1-c)(S-\mw^2)S\left( (1-c)^2\mw^4-(1+c)^2S^2 \right)  \nn\\
C_{21} =-C_{71} &=& \frac{1}{D_{W\gamma}}\sqrt{2(1-c^2)S}\mw\left( -(1-c)^3 \mw^6+(1-c)^2 (c+4) \mw^4 S \right.  \nn\\
&&\left.\hspace{8mm}+\left(c^3+c^2+3 c-5\right) \mw^2 S^2 +\left(c^3+c+2\right) S^3  \right)  \nn\\
C_{31} =-\sqrt{3}C_{81} &=& -\frac{1}{D_{W\gamma}}(1-c)S\left( 3 (1-c)^2 (c+1) \mw^6 +\left(6 c^2+c-7\right) \mw^4 S \right.  \nn\\
&&\left.\hspace{12mm} +\left(3 c^3+3 c^2+c+5\right) \mw^2 S^2 -(1-c) S^3 \right)  \nn\\
C_{32} = \frac{1}{2}C_{53} = \frac{1}{\sqrt{6}}C_{82}  &=& \frac{1}{D_{W\gamma}}(1-c)^2\sqrt{2(1-c^2)S}\mw\left( \mw^4S-2\mw^2S^2+S^3 \right)  \nn\\
C_{33} = -\frac{1}{2}C_{52} = \frac{1}{\sqrt{3}}C_{83} &=& -\frac{1}{2D_{W\gamma}}(1-c)(S-\mw^2)\left( -(1-c)^3 \mw^6+(1-c)^2 (c+3) \mw^4 S \right.  \nn\\
&&\left.\hspace{12mm} +\left(c^3+c^2+3 c-5\right) \mw^2 S^2+\left(c^3+c^2+3 c+3\right) S^3 \right)  \nn\\
C_{41} &=& \frac{1}{D_{W\gamma}}\left( (1-c)^4 \mw^8-4 (1-c)^3 \mw^6 S +2 \left(c^4+2 c^2-8 c+5\right) \mw^4 S^2 \right.  \nn\\
&&\left.\hspace{8mm} +4 \left(c^2+2 c-3\right) \mw^2 S^3 +\left(c^4+2 c^2+5\right) S^4 \right)  
\eear
}

Finally, the two eigenvalues of the reduced density matrix respect to the $\wpm$ boson are
\bear
\lambda_1^{red} &=& \frac{1}{2D_{W\gamma}}(1-c)^2 \left(S-\mw^2\right)^2 \left((1-c)^2 \mw^4+(c+1)^2 S^2\right)  \nn\\
\lambda_2^{red} &=& \frac{1}{2D_{W\gamma}} \left((1-c)^4 \mw^8 -2 (1-c)^4 \mw^6 S +2 (1-c)^2 \left(c^2+6\right) \mw^4 S^2 \right.  \nn\\
&&\left. \hspace{10mm}-2 \left(c^4-4 c^2-8 c+11\right) \mw^2 S^3 +\left(c^4+8 c^2+4 c+11\right) S^4  \right)
\label{WAtoWA-vNS}
\eear

The normalization of the reduced density matrix yields to $\lambda_1^{red}+\lambda_2^{red}=1$. Also, in the analyzed phase space, $D_{W\gamma}>0$ and it is easy to see that $\lambda_1^{red}\geq 0$ with equality if and only if $\cos(\theta)=1$ or $S=\mw^2$. In that kinematical configurations, $\lambda_2^{red}=1$ and the resulting Entropy of Entanglement of \eqref{definition-vNS} vanishes, then we conclude that the $\wpm\gamma$ final state is separable. This result was discussed in the text from the plot of the Negativity in the upper-left panel of \figref{qbqtplots}. 
The behaviour of the resulting $\vNS$ in the whole kinematical plane is shown in the upper-right panel of \figref{SEEremainingplots}

\subsection{$\gamma\gamma\to\ww$ process}

For this two-qutrits case, the decomposition of the 9$\times$9 density matrix is
\be
\rho = \frac{1}{9}I_{9} +\frac{1}{6}\sum_{i=1}^{8}A_i\lambda_i\otimes I_{3} +\frac{1}{6}\sum_{j=1}^{8}B_j I_{3}\otimes\lambda_j +\frac{1}{4}\sum_{i=1}^{8}\sum_{j=1}^{8}C_{ij} \lambda_i\otimes\lambda_j 
\ee

For a compact notation, we define the quantity
\be
D_{WW} = 48 \left(c^2-2\right)^2 \mw^4-8 \left(3 c^4-5 c^2+6\right) \mw^2 S+\left(3 c^4+2 c^2+11\right) S^2   
\ee

The resulting non-vanishing coefficients $A_i$ and $B_j$ are
{\allowdisplaybreaks
\bear
A_{2} = -A_7 = B_{2} = -B_7 &=& \frac{16}{D_{WW}} c \sqrt{2(1-c^2)} \mw S^{3/2}  \nn\\
A_{3} = -\sqrt{3}A_8 = B_{3} = -\sqrt{3}B_8 &=& \frac{4}{D_{WW}}S\left( 4 \left(2 c^2-3\right) \mw^2+\left(c^2+1\right) S \right)  \nn\\
A_{4} = B_4 &=& \frac{8}{D_{WW}}S\left( -4\mw^2 +(1+c^2)S \right)
\eear
}
and the non-vanishing coefficients $C_{ij}$ of the correlation matrix are
{\allowdisplaybreaks
\bear
C_{11} = C_{66} &=& -\frac{2}{D_{WW}}S\left( 4 \left(c^4-c^2+2\right) \mw^2 +(1-c^4)S \right)  \nn\\
C_{15} = C_{51} &=&  \nn\\
=-C_{56} = -C_{65} &=& -\frac{4}{D_{WW}}c\sqrt{2(1-c^2)S}\mw\left( 4 \left(c^2-2\right) \mw^2+(1-c^2)S \right)  \nn\\
C_{16} = C_{61} &=& \frac{4}{D_{WW}}\left( 8 \left(c^2-2\right)^2 \mw^4-2 \left(c^4-c^2+2\right) \mw^2 S-\left(1-c^2\right) S^2 \right)  \nn\\
C_{22} = C_{77} &=& \frac{2}{D_{WW}}S\left( \left(-12 c^4+12 c^2+8\right) \mw^2+(1-c^4)S \right)  \nn\\
C_{23} = C_{32} &=& \frac{8}{D_{WW}}c\sqrt{2(1-c^2)S}\mw\left( 2 \left(c^2-2\right) \mw^2+c^2 S \right)  \nn\\
C_{24} = C_{42} &=&  \nn\\
= -C_{47} = -C_{74} &=& \frac{4}{D_{WW}}c\sqrt{2(1-c^2)S}\mw\left( 4 \left(c^2-2\right) \mw^2+\left(c^2+3\right) S \right)  \nn\\
C_{27} = C_{72} &=& \frac{4}{D_{WW}}\left( 8 \left(c^2-2\right)^2 \mw^4+2 \left(3 c^4-3 c^2-2\right) \mw^2 S-\left(1-c^2\right) S^2 \right)  \nn\\
C_{28} = C_{82} &=& -\frac{8}{D_{WW}}c\sqrt{\frac{2}{3}(1-c^2)S}\mw\left( 6 \left(c^2-2\right) \mw^2+S \right)  \nn\\
C_{33} &=& \frac{2}{D_{WW}}\left( 8 \left(c^2-2\right)^2 \mw^4+4 \left(3 c^4-5 c^2+2\right) \mw^2 S +\left(c^4+1\right) S^2 \right)  \nn\\
C_{34} = C_{43} &=&  \nn\\
= -\sqrt{3}C_{48} = -\sqrt{3}C_{84} &=& \frac{4}{D_{WW}}S\left( \left(6 c^4-6 c^2-4\right) \mw^2 +\left(c^2+1\right) S \right)  \nn\\
C_{37} = C_{73} &=& -\frac{4}{D_{WW}}c\sqrt{2(1-c^2)S}\mw\left( 8 \left(c^2-2\right) \mw^2+\left(c^2+1\right) S \right)  \nn\\
C_{38} = C_{83} &=& \frac{4}{\sqrt{3}D_{WW}}\left( -12 \left(c^2-2\right)^2 \mw^4+2 \left(-3 c^4+c^2+6\right) \mw^2 S +\left(c^2-2\right) S^2 \right)  \nn\\
C_{44} &=& \frac{1}{D_{WW}}\left( 8 \left(2 \left(c^2-2\right) \mw^2+S\right)^2+2 \left(c^2+1\right)^2 S^2 \right)  \nn\\
C_{55} &=& \frac{1}{D_{WW}}\left( 8 \left(2 \left(c^2-2\right) \mw^2+S\right)^2-2 \left(c^2+1\right)^2 S^2 \right)  \nn\\
C_{78} = C_{87} &=& \frac{4}{D_{WW}}c(-1+3c^2)\sqrt{\frac{2}{3}(1-c^2)}\mw S^{3/2}  \nn\\
C_{88} &=& \frac{1}{3D_{WW}}\left( -48 \left(c^2-2\right)^2 \mw^4+8 \left(3 c^4-13 c^2+18\right) \mw^2 S \right.  \nn\\
&&\left.\hspace{12mm}+2 \left(3 c^4+4 c^2-5\right) S^2  \right)  
\eear
}

Finally, the three eigenvalues of the reduced density matrix that enter in \eqref{definition-vNS} are
\bear
\lambda_1^{red} &=& \frac{1}{D_{WW}}\left(4 \left(c^2-2\right) \mw^2+(1-c^2)S\right)^2  \nn\\
\lambda_2^{red} &=& \frac{1}{D_{WW}}\left(16 \left(c^2-2\right)^2 \mw^4-8 \left(c^4-c^2+2\right) \mw^2 S +(5+2c^2+c^4)S^2 \right. \nn\\  
&&\left.\hspace{8mm}-4S\sqrt{16 \left(c^2-2\right)^2 \mw^4 -8 \left(c^4-c^2+2\right) \mw^2 S +\left(c^2+1\right)^2 S^2} \right) \nn\\
\lambda_3^{red} &=& \frac{1}{D_{WW}}\left(16 \left(c^2-2\right)^2 \mw^4-8 \left(c^4-c^2+2\right) \mw^2 S +(5+2c^2+c^4)S^2 \right. \nn\\  
&&\left.\hspace{8mm}+4S\sqrt{16 \left(c^2-2\right)^2 \mw^4 -8 \left(c^4-c^2+2\right) \mw^2 S +\left(c^2+1\right)^2 S^2} \right)
\label{AAtoWW-vNS}
\eear
The behaviour of the resulting $\vNS$ in the whole kinematical plane is shown in the upper-left panel of \figref{SEEplots1}

\bibliography{Bell-VBS_arXiv2}

\end{document}